\documentclass[journal]{IEEEtran}

\usepackage{amsmath,amssymb,amsfonts}
\usepackage{epsfig,epstopdf,verbatim}
\usepackage{makecell}
\usepackage{algorithm,algpseudocode}
\usepackage{array}
\usepackage{stfloats}
\usepackage{graphicx}
\usepackage{cite,citesort}
\usepackage{ntheorem}
\usepackage{upgreek}
\usepackage{color}
\usepackage{bm,bbm}
\usepackage{graphics}
\usepackage[switch]{lineno}
\usepackage{threeparttable}
\usepackage{multicol,multirow}
\usepackage{soul}
\usepackage{diagbox}
\usepackage[caption=false,font=footnotesize]{subfig}
\usepackage{longtable}
\usepackage{xcolor}

\usepackage{tikz}
\usetikzlibrary{arrows.meta, positioning, calc}
\usepackage{amsmath,amssymb,bm}
\usetikzlibrary{shapes.geometric, arrows.meta, positioning, calc, fit, backgrounds}

\begin{document}

\title{Distributed Optimization-Learning with Graph Transformers for Terahertz Cell-Free Integrated Sensing and Communication Systems}
	
\author{\IEEEauthorblockN{Guangchen Wang, \IEEEmembership{Member, IEEE}, Zhifeng Tang, \IEEEmembership{Member, IEEE}, Nan Yang, \IEEEmembership{Senior Member, IEEE}, \\ Xin Hao, \IEEEmembership{Member, IEEE}, and Zhu Han, \IEEEmembership{Fellow, IEEE}}
\thanks{This work was supported by the Australian Research Council Discovery Project (DP230100878).}
\thanks{G. Wang, Z. Tang, and N. Yang are with the School of Engineering, Australian National University, Canberra, ACT 2601, Australia (E-mails: \{guangchen.wang, zhifeng.tang, nan.yang\}@anu.edu.au).}\vspace{-20pt}
\thanks{X. Hao is with the Faculty of Engineering and Information Technology, University of Technology Sydney, Sydney, NSW 2007, Australia (E-mail: haoxin1022@hotmail.com).}
\thanks{Z. Han is with the Department of Electrical and Computer Engineering, University of Houston, Houston, TX 77004, USA (E-mail: hanzhu22@gmail.com).}
\vspace{-20pt}}

\markboth{Submitted to IEEE Transactions on Wireless Communications}%
{Wang \MakeLowercase{\textit{et al.}}: Distributed Optimization-Learning with Graph Transformers for Terahertz Cell-Free Integrated Sensing and Communication Systems}

\maketitle

\begin{abstract}
In this paper, we propose a distributed optimization-learning framework for terahertz (THz) cell-free integrated sensing and communication (CF-ISAC) systems, termed Distributed Optimization-Learning with Graph Transformers (DOLG). We first formulate a highly non-convex joint scheduling and signal design problem for THz CF-ISAC systems, jointly optimizing access point (AP)-user equipment (UE) association and beamforming under signal to interference plus noise ratio based communication and Cram\'{e}r-Rao bound based sensing constraints, together with line-of-sight-driven visibility rules and per-AP power constraints. We also develop an optimization based benchmark utilizing a tractable relaxed reformulation. Building upon this optimization structure, we redesign a graph transformer network (GTN) as an optimization-aware representation module that encodes cross-field wavefront geometry, blockage visibility, and sensing relevance in a permutation-equivariant manner. The proposed DOLG framework amortizes the iterative optimization procedure into a scalable GTN-conditioned distributed multi-agent reinforcement learning policy through centralized training and decentralized execution, while preserving per-AP power constraints via structure-preserving projections. Simulation results demonstrate that the proposed DOLG framework achieves stable convergence and effectively balances the communication–sensing tradeoff. From the system-level perspective, it outperforms multicell and non-joint design baselines. Furthermore, it surpasses conventional optimization based and heuristic approaches in terms of both ISAC performance and computational scalability.
\end{abstract}


\section{Introduction}\label{sec:I_introduction}

\IEEEPARstart{T}{he rapid} proliferation of large-scale intelligent systems, including mobile sensing systems, distributed robotic platforms, smart grids, and collaborative machine learning, has intensified the demand for distributed optimization and inference under stringent communication constraints \cite{9446488}. In such systems, a large number of geographically distributed agents must cooperatively perform complex tasks, such as estimating global statistics, optimizing shared objectives, or forming consistent probabilistic beliefs, based solely on local information exchanges \cite{4663899,10.1145/3377454}. However, the performance of distributed cooperation is fundamentally constrained by the underlying communication graph, as limited bandwidth, imperfect quantization, unreliable links, and heterogeneous connectivity jointly restrict information exchange and degrade overall system performance \cite{xiao2004fast}.

Driven by the increasing density and scale of emerging systems, distributed learning and inference have become pivotal research paradigms. Early studies primarily addressed on average consensus and decentralized optimization under idealized or static connectivity \cite{4118472,8664630,1431045}. More recently, research has shifted toward communication-efficient aggregation, model compression, and robust information fusion to mitigate bandwidth and reliability bottlenecks \cite{8714026,9530694,10038471}. Despite these advancements, most existing cooperation schemes remain tailored to low-frequency wireless or wired systems. Therefore, they are not directly applicable to future intelligent infrastructures, which are characterized by the simultaneous presence of high mobility, pronounced spatial heterogeneity, and stringent communication-sensing requirements.

Meanwhile, the sixth-generation (6G) wireless systems are envisioned to leverage terahertz (THz) communications to support integrated sensing and communication (ISAC) using cell-free (CF) architectures \cite{9737357,9933498,10552627}, which fundamentally complicate distributed cooperation. In particular, THz communications suffers from severe molecular absorption, extremely directional beamforming, and pronounced near-field effects due to ultra-short wavelengths \cite{10494372}. The resulting coexistence of near- and far-field propagation induces wavefront distortions and highly nonlinear channel responses \cite{10616028}, thus rendering conventional distributed protocols inefficient. Moreover, dynamic blockages and time-varying line-of-sight (LoS) conditions frequently disrupt network connectivity \cite{9247469,11007489}, posing additional challenges to reliable information exchange in distributed optimization. This complexity is further amplified in CF-ISAC systems. Specifically, a large number of distributed access points (APs) jointly serve user equipments (UEs) and sensing targets (STs) without predefined cell boundaries, where each AP must coordinate sensing and communication signals and optimize resource allocation based solely on partial local observations. Consequently, system-wide communication-sensing performance critically relies on efficient distributed cooperation among APs, which is particularly difficult to achieve under dynamically evolving cross-field topologies \cite{7827017,9064545,10742291}. These challenges are further amplified in cross near- and far-field THz propagation environments, where the cross-field channel representation induces strong coupling among AP, UE and ST locations, and array responses, resulting in a highly non-convex joint scheduling and signal design problem \cite{11062661}.

Classical solution approaches, including convex approximation, branch-and-bound, and relaxation based optimization, are capable of handling small-scale problem instances \cite{5447076,lawler1966branch,6853715}. However, their prohibitive computational complexity, limited scalability, and poor adaptability to rapidly changing system states limit their practical applicability. Recent learning based methods, such as deep reinforcement learning (DRL) and graph neural networks (GNNs), have shown the potential to address sequential decision-making problems in the aforementioned complex wireless environments \cite{8227766,10122232,10161704}. Nevertheless, DRL based approaches typically lack explicit awareness of the underlying optimization structure, and therefore cannot guarantee constraint satisfaction, while conventional GNNs \cite{Xin_HML} are insufficient to fully capture cross-field wavefront geometry and blockage-induced graph dynamics.

To address these challenges, we propose Distributed Optimization-Learning with Graph Transformers (DOLG), an optimization-aware distributed control framework for THz CF-ISAC systems where multiple distributed APs collaboratively serve multiple UEs and STs in a cross near-/far-field THz environment under LoS blockage and communication-sensing coupling. Leveraging graph transformer networks (GTNs), DOLG integrates graph based structured representation, optimization-aligned decision design, and distributed policy learning to capture the heterogeneous interactions among APs, UEs, and STs under cross near-/far-field propagation, LoS-driven blockages, and communication-sensing coupling. To reveal the problem's underlying structure, we further develop an optimization based benchmark obtained from a tractable relaxed reformulation of the original problem. Guided by this benchmark, DOLG converts the original high-complexity design problem into a scalable distributed learning framework, enabling efficient online ISAC decision-making while preserving strong alignment with the underlying communication-sensing optimization problem. In summary, our main contributions are summarized as follows:
\begin{itemize}
\item We develop a unified THz CF-ISAC system model that captures cross-field wavefront characteristics, molecular absorption, blockage effects, and sensing-driven position refinement, enabling accurate THz propagation modeling.
\item We formulate a joint scheduling and signal design problem for the THz CF-ISAC system, where AP-UE association and beamforming are jointly optimized under signal to interference plus noise ratio (SINR) based communication constraints, Cram\'{e}r-Rao bound (CRB) based sensing requirements, LoS-driven visibility rules, and per-AP power constraints.
\item We develop an optimization-based benchmark to reveal the underlying problem structure. Specifically, we transform the formulated problem into a tractable relaxed formulation, where block coordinate descent (BCD), successive convex approximation (SCA), and semidefinite relaxation (SDR) are jointly integrated into a unified optimization architecture, termed BCD-SCA-SDR (B2S).
\item We design the GTN as an optimization-aware representation module that captures cross-field wavefront geometry, blockage visibility, multi-AP coupling, and sensing relevance in a permutation-equivariant manner. This design interfaces with the optimization benchmark through warm-start initialization and adaptive feasibility control.
\item We propose DOLG, a GTN-conditioned distributed optimization-learning framework under centralized training and decentralized execution. By modeling each AP as an autonomous agent within a multi-agent reinforcement learning (MARL) policy, DOLG converts the B2S-inspired optimization structure into a scalable distributed policy, ensuring per-AP power feasibility through optimization-aligned action design and structure-preserving feasibility projections.
\item Simulation results demonstrate that DOLG achieves stable convergence and effectively balances the communication–sensing tradeoff. We show that DOLG outperforms multicell and non-joint design baselines from a system-design perspective, while surpassing conventional optimization based and heuristic methods in terms of both ISAC performance and computational scalability.
\end{itemize}

The remainder of this paper is organized as follows. Section~\ref{sec:II_system} introduces the THz CF-ISAC system model, including the system topology, THz channel model, ISAC signal model, and problem formulation. Section~\ref{sec:opt_solution} presents the problem transformation and develops the optimization based B2S solution. Section~\ref{sec:GTN_main} redesigns the graph transformer network as an optimization-aware structured representation module. Section~\ref{sec:DOLG} proposes the distributed optimization-learning framework with graph transformers for the scalable CF-ISAC system. Section~\ref{sec:simulation_results} provides numerical results to verify the effectiveness of the proposed framework, and Section~\ref{sec:VII_conclusion} concludes the paper.

Notations: Matrices, vectors, scalars, and sets are denoted by uppercase boldface, lowercase boldface, italic, and calligraphic symbols, respectively. For a matrix $\mathbf{A}$, $\mathbf{A}^{\mathsf T}$, $\mathbf{A}^{\mathsf H}$, and $\mathbf{A}^{-1}$ denote its transpose, Hermitian transpose, and inverse, respectively. The operators $\operatorname{tr}(\cdot)$, $\det(\cdot)$, $\mathrm{blkdiag}(\cdot)$, and $\mathrm{vec}(\cdot)$ denote the trace, determinant, block-diagonalization, and vectorization, respectively. In addition, $|\cdot|$, $\|\cdot\|$, $\mathbb{E}[\cdot]$, $\operatorname{Re}\{\cdot\}$, and $\operatorname{Im}\{\cdot\}$ denote the magnitude of a complex scalar, Euclidean norm, expectation, real part, and imaginary part, respectively. The symbols $\mathbb{R}$ and $\mathbb{C}$ denote the real and complex fields, respectively.

\section{System Model}\label{sec:II_system}

\subsection{System Topology}\label{sec:II_system_A}

\begin{figure}[!t]
    \centering
    \centerline{\includegraphics[width=\columnwidth]{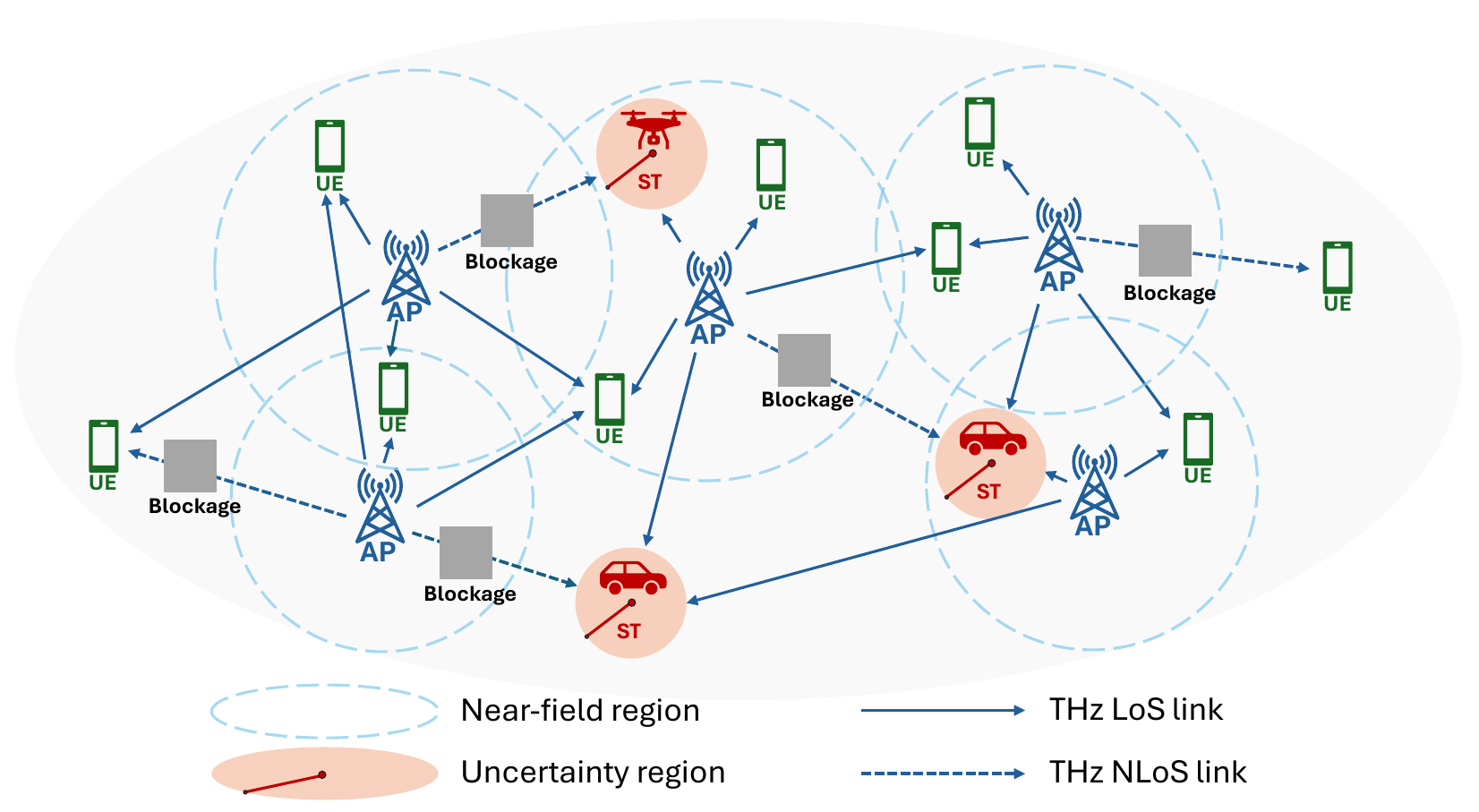}}
    \caption{Illustration of the THz CF-ISAC system with cross-field propagation.}
    \label{fig:sys}
\end{figure}

We consider the downlink of a THz CF-ISAC system, as depicted in Fig. \ref{fig:sys}, where $M$ distributed APs, indexed by $\mathcal{M}=\{1,\dots,M\}$, collaboratively serve $K$ communication UEs, indexed by $\mathcal{K}=\{1,\dots,K\}$, and $S$ STs, indexed by $\mathcal{S}=\{1,\dots,S\}$, in a CF manner. To establish a unified modeling framework, we introduce a unified generic entity (GE) index set $\mathcal{X}=\{1,\ldots,K+S\}$, which is partitioned into $\mathcal{X}_{\mathrm{UE}}=\mathcal{K}$ and $\mathcal{X}_{\mathrm{ST}}=\{K+1,\ldots,K+S\}$. Specifically, each ST index, $s\in\mathcal{S}$, is mapped to $x=K+s\in\mathcal{X}_{\mathrm{ST}}$. Therefore, a generic index $x\in\mathcal{X}$ refers to a UE if $x\in\mathcal{X}_{\mathrm{UE}}$, and an ST if $x\in\mathcal{X}_{\mathrm{ST}}$. 

In our considered system, each AP is assumed to be equipped with a uniform linear array (ULA) consisting of $N_m$ antennas, while all UEs are assumed to be single-antenna receivers \cite{7827017,9064545}. We denote $\mathbf{p}_m\in\mathbb{R}^2$, $\mathbf{u}_k\in\mathbb{R}^2$, and $\mathbf{q}_s\in\mathbb{R}^2$ as the locations of the $m$-th AP, the $k$-th UE, and the $s$-th ST, respectively. Due to channel dynamics and sensing uncertainty, we assume that the $s$-th ST is initially located within a circular region, given by
\begin{equation}
\mathcal{C}_s = \{\mathbf{q}_{s}\in\mathbb{R}^2:\|\mathbf{q}_{s}-\bar{\mathbf{q}}_s\|\le\rho_s\},
\end{equation}
where $\bar{\mathbf{q}}_s$ and $\rho_s$ denote the center and radius of the circular region of uncertainty, respectively.

For the link between the $m$-th AP and the $x$-th GE, we define the displacement vector $\mathbf{r}_{m,x}$ and the corresponding link distance $r_{m,x}\triangleq\|\mathbf{r}_{m,x}\|$. We note that the Rayleigh distance, given by $R_m = \frac{2D_m^2}{\lambda}$, is typically employed to distinguish the far-field and near-field propagation regimes~\cite{11194115,9693928}, where $D_m$ is the antenna aperture size of the $m$-th AP and $\lambda$ is the wavelength. Accordingly, a UE may be simultaneously located in the near-field of some APs and in the far-field of others, leading to a cross-field system topology.

\subsection{THz Channel Model}\label{sec:II_system_B}

\subsubsection{Pathloss and Molecular Absorption}

For the link between the $m$-th AP and the $x$-th GE, the baseband pathloss accounting for free-space spreading and molecular absorption is expressed as
\begin{equation} \label{eq:THz_pathloss}
L_{m,x}=\left(\frac{4\pi f r_{m,x}}{c}\right)^{2}e^{\kappa(f)r_{m,x}},
\end{equation}
where $c=3\times10^8$ m/s is the speed of light, $f$ is the operating frequency, and $\kappa(f)$ is the molecular absorption coefficient at the frequency $f$. The corresponding complex gain is expressed as $\beta_{m,x}=\sqrt{\frac{G_t G_r}{L_{m,x}}}$, where $G_t$ and $G_r$ represent the gain of each individual antenna element at the AP and GE, respectively. For STs, which are modeled as passive point scatterers, we set $G_r=1$.

\subsubsection{Array Response Models}

When $r_{m,x}<R_m$, the wavefront exhibits noticeable spherical curvature \cite{7981398}, corresponding to the near-field regime. As a result, the channel steering element between the $n$-th antenna element of the $m$-th AP and the $x$-th GE is expressed as
\begin{equation}
a^{(n)}(\theta_{m,x},r_{m,x})=e^{-j\tfrac{2\pi f}{c}\,r_{m,x,n}},
\end{equation}
where $r_{m,x,n}=\|\mathbf{r}_{m,x}-\mathbf{d}_{m,n}\|$ is the distance between the $n$-th antenna element of the $m$-th AP and the $x$-th GE, with $\{\mathbf{d}_{m,n}\}_{n=1}^{N_m}$ representing the coordinates of the $n$-th antenna element of the $m$-th AP. 

When $r_{m,x}\geq R_m$, the wavefront can be approximated by the planar model, corresponding to the far-field regime, where the channel steering element between the $n$-th antenna element of the $m$-th AP and the $x$-th GE is approximated by
\begin{equation}\label{eq:far_field_response}
a^{(n)}(\theta_{m,x},r_{m,x})\approx e^{-j\tfrac{2\pi}{\lambda}\left(r_{m,x}-\mathbf{u}^{\mathsf T}(\theta_{m,x})\mathbf{d}_{m,n}\right)},
\end{equation}
where $\mathbf{u}(\theta_{m,x})$ denotes the unit-direction vector associated with angle $\theta_{m,x}$. By stacking all $N_m$ channel steering elements, we obtain the unified channel steering vector as
\begin{equation}\label{eq:unified_response}
\mathbf{a}(\theta_{m,x},r_{m,x})=\left[a^{(n)}(\theta_{m,x},r_{m,x})\right]_{n=1}^{N_m}.
\end{equation}
For the downlink transmission from the $m$-th AP to the $x$-th GE, the complex channel vector is expressed as
\begin{equation}\label{eq:downlink_channel}
\mathbf{h}_{m,x}=\beta_{m,x} \mathbf{a}(\theta_{m,x}, r_{m,x}) \in \mathbb{C}^{N_m \times 1},
\end{equation}
which captures THz-specific propagation phenomena, including pathloss and molecular absorption, and incorporates both near- and far-field effects in the array response.

\subsection{ISAC Signal Model}\label{sec:II_system_C}

\subsubsection{Downlink Joint Transmission}

Each AP transmits a composite ISAC waveform combining data and sensing pilots. The transmitted signal from the $m$-th AP at time slot $t$ is expressed as
\begin{equation}\label{eq:transmitted_signal} 
\mathbf{x}_m[t]=\sum_{k=1}^{K}\delta_{m,k}\mathbf{w}_{m,k}[t]\,c_k[t]+\sum_{s=1}^{S}\delta_{m,s}\mathbf{s}_{s}[t],
\end{equation}
where $\mathbf{w}_{m,k}[t]\in \mathbb{C}^{N_m\times1}$ is the precoding vector, $c_k[t]$ is the data symbol for the $k$-th UE, and $\mathbf{s}_s[t]$ is the dedicated sensing signal. Here, $\delta_{m,k}\in\{0,1\}$ indicates whether the $m$-th AP serves the $k$-th UE, while $\delta_{m,s}\in\{0,1\}$ indicates whether the $m$-th AP participates in sensing the $s$-th ST.

Based on $\mathbf{h}_{m,k}$ in \eqref{eq:downlink_channel} and $\mathbf{x}_m[t]$ in \eqref{eq:transmitted_signal}, the received signal at the $k$-th UE at time slot $t$ is given by
\begin{align}\label{eq:yk_receive}
y_k[t]
&=\sum_{m=1}^{M}\mathbf{h}_{m,k}^{\mathsf H}[t]\mathbf{x}_m[t]
=\sum_{m=1}^{M}\delta_{m,k}\mathbf{h}_{m,k}^{\mathsf H}[t]\mathbf{w}_{m,k}[t]c_k[t]\notag\\
&\quad+\sum_{j\neq k}\sum_{m=1}^{M}\delta_{m,j}\mathbf{h}_{m,k}^{\mathsf H}[t]\mathbf{w}_{m,j}[t]c_j[t]\notag\\
&\quad+\sum_{s=1}^{S}\sum_{m=1}^{M}\delta_{m,s}\mathbf{h}_{m,k}^{\mathsf H}[t]\mathbf{s}_{s}[t]+z_k[t],
\end{align}
where the first term represents the coherently combined desired signal, the second term accounts for multiuser interference, the third term captures the structured interference introduced by sensing pilots, and $z_k[t]\sim \mathcal{CN}(0,\sigma^2)$ denotes the additive white Gaussian noise (AWGN) at the $k$-th UE. To characterize the communication quality, we calculate the instantaneous SINR at the $k$-th UE as
\begin{equation}\label{eq:SINR_k}
\gamma_k[t] =
\frac{1}{I_k[t] + \sigma^2}
\left|
\displaystyle\sum_{m=1}^{M}
\delta_{m,k}
\mathbf{h}_{m,k}^{\mathsf H}[t]\mathbf{w}_{m,k}[t]
\right|^2,
\end{equation}
where $I_k [t]$ is the corresponding aggregated interference, given by
{\begin{align}\label{eq:Ik_def}
I_k [t] \triangleq
\sum_{j\neq k}
\Bigg|
\sum_{m=1}^{M}
&\delta_{m,j}
\,\mathbf{h}_{m,k}^{\mathsf H}[t]
\mathbf{w}_{m,j}[t]
\Bigg|^{2} \notag\\
&\quad+
\sum_{s=1}^{S}
\Bigg|
\sum_{m=1}^{M}
\delta_{m,s}
\,\mathbf{h}_{m,k}^{\mathsf H}[t]
\mathbf{s}_{s}[t]
\Bigg|^{2}.
\end{align}}
We clarify that the SINR expression in \eqref{eq:SINR_k} reveals the inherent coupling between data transmission and sensing-induced interference, as both data precoders and sensing pilots propagate through the unified cross-field THz channel.

\subsubsection{Sensing Echo Model}

The unified ISAC waveform transmitted by APs is also exploited to sense the locations of STs \cite{9540344}. During each coherence block, the embedded sensing pilots illuminate the STs, and the reflected echoes are captured at the APs. For the $m$-th AP and the $s$-th ST, the near-field round-trip channel matrix is given by
\begin{align}
\mathbf{G}_{m,s} = \beta_{m,s}^{\mathrm{rt}} \mathbf{a}(\theta_{m,s},r_{m,s})\mathbf{a}^{\mathsf T}(\theta_{m,s},r_{m,s}),
\end{align}
where $\theta_{m,s}$ and $r_{m,s}$ denote the angle and distance from the $m$-th AP to the $s$-th ST, and $\beta_{m,s}^{\mathrm{rt}} \triangleq \beta_{m,s}^2$  represents the round-trip complex gain. Then, the echo signal received at the $m$-th AP is given by
\begin{align}
\mathbf{y}_{m,s}^{\mathrm{echo}}[t] = \mathbf{G}_{m,s}\mathbf{x}_{m}[t] + \mathbf{z}_{m,s}[t],
\end{align}
where $\mathbf{z}_{m,s}[t]$ denotes sensing noise at the $m$-th AP. Stacking the echoes received at all APs yields $\mathbf{y}_{s}^{\mathrm{echo}}[t]=\big[(\mathbf{y}_{1,s}^{\mathrm{echo}}[t])^{\mathsf T},\ldots,(\mathbf{y}_{M,s}^{\mathrm{echo}}[t])^{\mathsf T}\big]^{\mathsf T}.$ The objective of sensing is to estimate the parameters of STs from the received echo signal samples over the entire coherence time block, i.e., $\mathbf{Y}_s^{\mathrm{echo}} = [\mathbf{y}_s^{\mathrm{echo}}(1), \mathbf{y}_s^{\mathrm{echo}}(2), \cdots, \mathbf{y}_s^{\mathrm{echo}}(T)]$. We employ the CRB to characterize the sensing performance \cite{kay1993fundamentals}, providing a lower bound on the mean squared error (MSE) of an unbiased estimator, which is given by
\begin{align}\label{eq:crb}
\textrm{CRB}(\mathbb{X},\mathbf{G}, N_0) = (\mathbf{J}_{11}-\mathbf{J}_{12}\mathbf{J}_{22}^{-1}\mathbf{J}_{12}^{\mathsf H})^{-1},
\end{align}
where $N_0$ denotes the noise power, and $\mathbf{J}_{11}$, $\mathbf{J}_{12}$, and $\mathbf{J}_{22}$ are the corresponding submatrices of the Fisher information matrix (FIM) with respect to the unknown parameters \cite{8828023}. For the multi-AP sensing channel, the composite unknown parameter vector for the $s$-th ST is defined as $\boldsymbol{\xi}_s=\left[\boldsymbol{\xi}_{1,s}^{\mathsf T},\ldots,\boldsymbol{\xi}_{M,s}^{\mathsf T}\right]^{\mathsf T}$, where each per-AP parameter subvector is given by $\boldsymbol{\xi}_{m,s}=\left[r_{m,s},\theta_{m,s},\operatorname{Re}\{\beta_{m,s}^{\mathrm{rt}}\},\operatorname{Im}\{\beta_{m,s}^{\mathrm{rt}}\}\right]^{\mathsf T}$. We denote the vectorized received echo signal by $\mathbf{Y}_s^{\mathrm{vec}}=\mathrm{vec}(\mathbf{Y}_s^{\mathrm{echo}})$, and its mean vector by $\boldsymbol{\Upsilon}(\boldsymbol{\xi}_s)\triangleq\mathbb{E}[\mathbf{Y}_s^{\mathrm{vec}}]$. Since the sensing noise follows a circularly symmetric complex Gaussian distribution, we have $\mathbf{Y}_s^{\mathrm{vec}} \sim \mathcal{CN}(\boldsymbol{\Upsilon}(\boldsymbol{\xi}_s),N_0\mathbf{I})$, where $\mathbf{I}$ denotes the identity matrix with appropriate dimensions. For complex Gaussian observations whose mean depends only on $\boldsymbol{\xi}_s$, the FIM is calculated by
\begin{align} \label{eq:FIM_main}
\mathbf{J}_{\boldsymbol{\xi}_s} 
= \frac{2}{N_0}\operatorname{Re}\left\{
\left(
\frac{\partial \boldsymbol{\Upsilon}}
{\partial \boldsymbol{\xi}_s}
\right)^{\mathsf H}
\left(
\frac{\partial \boldsymbol{\Upsilon}}
{\partial \boldsymbol{\xi}_s}
\right)
\right\}.
\end{align}

Due to the independence of sensing noises across APs, $\mathbf{J}_{\boldsymbol{\xi}_s}$ admits a block-diagonal structure, i.e., $\mathbf{J}_{\boldsymbol{\xi}_s}=\mathrm{blkdiag}\{\mathbf{J}_{\boldsymbol{\xi}_{1,s}},\ldots,\mathbf{J}_{\boldsymbol{\xi}_{M,s}}\}$. Accordingly, each per-AP FIM block, $\mathbf{J}_{\boldsymbol{\xi}_{m,s}}\in\mathbb{C}^{4\times4}$, is partitioned as
\begin{align} \label{eq:FIM_block_m}
\mathbf{J}_{\boldsymbol{\xi}_{m,s}}
=
\left[
\begin{aligned}
&\mathbf{J}_{11,m,s} &\mathbf{J}_{12,m,s}\\
&\mathbf{J}_{12,m,s}^{\mathsf H} &\mathbf{J}_{22,m,s}
\end{aligned}
\right],
\end{align}
where $\mathbf{J}_{11,m,s}$, $\mathbf{J}_{12,m,s}$, and $\mathbf{J}_{22,m,s}$ are obtained as
\begin{align} \label{eq:FIM_J11}
\mathbf{J}_{11,m,s} =\left[
\begin{aligned}
&J_{r_{m,s}, r_{m,s}} &J_{r_{m,s}, \theta_{m,s}}\\
&J_{r_{m,s}, \theta_{m,s}} &J_{\theta_{m,s}, \theta_{m,s}}
\end{aligned}
\right],
\end{align}
\begin{align} \label{eq:FIM_J12}
\mathbf{J}_{12,m,s} =\left[
\begin{aligned}
&J_{r_{m,s}, \operatorname{Re}(\beta_{m,s}^{\mathrm{rt}})} 
&J_{r_{m,s}, \operatorname{Im}(\beta_{m,s}^{\mathrm{rt}})}\\
&J_{\theta_{m,s}, \operatorname{Re}(\beta_{m,s}^{\mathrm{rt}})} 
&J_{\theta_{m,s}, \operatorname{Im}(\beta_{m,s}^{\mathrm{rt}})}
\end{aligned}
\right],
\end{align}
and
\begin{align} \label{eq:FIM_J22}
\mathbf{J}_{22,m,s}=\left[
\begin{aligned}
&J_{\operatorname{Re}(\beta_{m,s}^{\mathrm{rt}}),
\operatorname{Re}(\beta_{m,s}^{\mathrm{rt}})} &0\\
&0 &J_{\operatorname{Im}(\beta_{m,s}^{\mathrm{rt}}),
\operatorname{Im}(\beta_{m,s}^{\mathrm{rt}})}
\end{aligned}
\right],
\end{align}
respectively. The detailed derivation of \eqref{eq:FIM_main}--\eqref{eq:FIM_J22}, including the Jacobians and trace-form expressions, is provided in Appendix~A.
\subsection{Cooperative ISAC Problem Formulation}\label{sec:II_system_D}

We assume that each UE selects a LoS-driven candidate subset of APs that participate in joint communication-sensing transmission, where APs whose LoS probability exceeds a threshold $p_{\mathrm{th}}$ are included, i.e.,
\begin{equation}\label{eq:LOS_AP_set}
\mathcal{M}_k=\left\{ m \in \mathcal{M} \,\middle|\, p_{\mathrm{LoS}}(r_{m,k}) \geq p_{\mathrm{th}} \right\}.
\end{equation}
This candidate set induces a fixed LoS-driven visibility mask $\xi_{m,k}\in\{0,1\}$. In this work, we optimize the AP-UE association coefficients $\{\delta_{m,k}\}$ only. The AP-ST sensing participation indicators $\{\delta_{m,s}\}$ are pre-determined by visibility and kept fixed during optimization. Accordingly, the binary AP association coefficient $\delta_{m,k}\in\{0,1\}$ is optimized subject to $\delta_{m,k}\leq \xi_{m,k}$. In addition, the received SINR at each UE must satisfy the SINR constraint, given by $\gamma_k \ge \gamma_{\mathrm{th}}$, $\forall k \in \mathcal{K}$, where $\gamma_{\mathrm{th}}$ is the target SINR threshold. Here, $\gamma_k$ is an implicit function of the beamforming variables and AP-UE association coefficients, and the time-slot index is omitted for notational simplicity.

In the CF architecture, all APs collaboratively serve all visible UEs without user-centric clustering under a per-AP transmit power constraint $P_{\mathrm{max}}$, which is given by $\sum_{k\in\mathcal{K}}\|\mathbf{w}_{m,k}\|^2 \leq P_{\mathrm{max}}$, $m\in\mathcal{M}$.
For sensing, the system estimates the parameters associated with a set of STs, $\mathcal{S}$. The sensing accuracy of each ST, $s\in\mathcal{S}$, is characterized by the CRB, denoted by $\mathrm{CRB}_s(\{\mathbf{w}_{m,k}\})$. The accuracy requirement imposed by the $s$-th ST is formulated as $\mathrm{CRB}_{s}(\{\mathbf{w}_{m,k}\})\leq\epsilon_{\mathrm{th}}$, $\forall s \in \mathcal{S}$, where $\epsilon_{\mathrm{th}}$ is the CRB threshold, which guarantees reliable detection and localization performance.

By jointly designing the transmit beamformers, $\{\mathbf{w}_{m,k}\}$, and the AP-UE association coefficients, $\{\delta_{m,k}\}$, the system seeks to minimize the total sensing error while guaranteeing minimum communication quality for each UE, sensing accuracy for each ST, and enforcing per-AP power constraints. The resulting optimization problem is formulated as
\begin{subequations}\label{prob:P0}
\begin{align}
\mathcal{P}_0:&\quad\min_{\{\mathbf{w}_{m,k},\delta_{m,k}\}}\sum_{s=1}^{S}\mathrm{CRB}_{s}(\{\mathbf{w}_{m,k}\})
\label{prob:P0a} \\
\mathrm{s.t.}
&\quad\gamma_k \geq \gamma_{\mathrm{th}},
\quad \forall k\in\mathcal{K},
\label{prob:P0b} \\
& \quad \sum_{k\in\mathcal{K}}
\|\mathbf{w}_{m,k}\|^2
\leq P_{\mathrm{max}},
\ \ \forall m\in\mathcal{M},
\label{prob:P0c} \\
& \quad \mathrm{CRB}_{s}(\{\mathbf{w}_{m,k}\})
\leq \epsilon_{\mathrm{th}},
\ \ \forall s\in\mathcal{S},
\label{prob:P0d} \\
& \quad \delta_{m,k}\in\{0,1\},
\ \  \forall m\in\mathcal{M},\ k\in\mathcal{K}, \\
& \quad \delta_{m,k}\le \xi_{m,k},
\quad \forall m\in\mathcal{M},\ k\in\mathcal{K}.
\label{prob:P0e}
\end{align}
\end{subequations}

\section{Problem Transformation and Optimization Based Solution}\label{sec:opt_solution}

To enable tractable algorithm design for the highly non-convex mixed-integer ISAC optimization problem $\mathcal{P}_0$, we first apply integer relaxation and derive a sequence of tractable reformulations and surrogates \cite{7547360}. Building upon these transformations, we develop a three-stage BCD-SCA-SDR algorithm, termed B2S, to successively solve convex subproblems and converge to a stationary point of the relaxed formulation.
\subsection{Problem Analysis}

We first analyze the joint ISAC optimization problem in \eqref{prob:P0} for the cross-field THz CF-ISAC system. Owing to the coupling among continuous beamforming variables and discrete association variables under the cross-field model, the problem constitutes a mixed-integer nonlinear programming (MINLP) problem, which is inherently non-convex. In particular, the combinatorial association decisions and the non-convex SINR and CRB expressions induced by cross-field propagation render the problem intractable for direct optimization.

\subsubsection{Association-Induced Combinatorial Non-Convexity}

The joint ISAC design involves continuous beamforming variables $\mathbf{w}_{m,k}$ and discrete AP association coefficients $\delta_{m,k}\in\{0,1\}$. While the LoS-driven visibility mask, $\xi_{m,k}$, determines the candidate AP set for each UE, the binary variable $\delta_{m,k}$ specifies whether an LoS-eligible AP is selected for transmission. Through effective channel construction, these association variables couple with both the communication SINR and sensing CRB expressions. Consequently, the resulting optimization problem is a MINLP problem with an inherently non-convex feasible set, due to the combinatorial association decisions.

\subsubsection{Non-Convexity of SINR Constraints}

The instantaneous SINR at the $k$-th $\mathrm{UE}$, introduced in \eqref{eq:SINR_k}--\eqref{eq:Ik_def}, is a fractional quadratic function of both downlink communication beamformers and sensing waveforms. We define the stacked multi-AP channel vector and beamforming vector as
\begin{align}
\tilde{\mathbf{h}}_k^{\mathsf H}[t]
\triangleq
\left[
\mathbf{h}_{1,k}^{\mathsf H}[t],\;
\ldots,\;
\mathbf{h}_{M,k}^{\mathsf H}[t]
\right]
\in\mathbb{C}^{1\times N_{\rm tot}}
\end{align}
and
\begin{align}
\tilde{\mathbf{w}}_k[t]
\triangleq
\left[
\mathbf{w}_{1,k}^{\mathsf T}[t],\;
\ldots,\;
\mathbf{w}_{M,k}^{\mathsf T}[t]
\right]^{\mathsf T}
\in\mathbb{C}^{N_{\rm tot}\times 1},
\end{align}
respectively, where $N_{\rm tot}=\sum_{m=1}^{M}N_m$ is the total number of antennas across all APs. Moreover, we define the association-selection matrix as
\begin{align}
\mathbf{D}_k
\triangleq
\mathrm{blkdiag}\big(\delta_{1,k}\mathbf{I}_{N_1},\ldots,\delta_{M,k}\mathbf{I}_{N_M}\big).
\end{align}
Based on such defined notations, the desired signal term in the numerator of \eqref{eq:SINR_k} is obtained as
\begin{align}\label{eq:signal_quadratic_form}
\left|
\sum_{m=1}^{M}
\delta_{m,k}\,
\mathbf{h}_{m,k}^{\mathsf H}[t]\mathbf{w}_{m,k}[t]
\right|^2
=\tilde{\mathbf{w}}_k^{\mathsf H}[t]\,
\mathbf{D}_k\mathbf{H}_k[t]\mathbf{D}_k\,
\tilde{\mathbf{w}}_k[t],
\end{align}
where $\mathbf{H}_k[t] \triangleq \tilde{\mathbf{h}}_k[t]\tilde{\mathbf{h}}_k^{\mathsf H}[t] \succeq \mathbf{0}$.
Similarly, the first term of $I_k[t]$ in \eqref{eq:Ik_def} is obtained as
\begin{align}\label{eq:interference_quadratic_form}
\sum_{j\neq k}
\left|\sum_{m=1}^{M}\delta_{m,j}\,
\mathbf{h}_{m,k}^{\mathsf H}[t]\mathbf{w}_{m,j}[t]\right|^2 
=\sum_{j\neq k}\tilde{\mathbf{w}}_j^{\mathsf H}[t]\,\mathbf{D}_j\mathbf{H}_k[t]\mathbf{D}_j\,\tilde{\mathbf{w}}_j[t],
\end{align}
which is a sum of quadratic forms in the precoders, $\tilde{\mathbf{w}}_j[t]$. Let $I_k^{\mathrm{MU}}[t]$ and $I_k^{\mathrm{SI}}[t]$ denote the multiuser interference and sensing-induced interference, respectively, such that $I_k[t]=I_k^{\mathrm{MU}}[t]+I_k^{\mathrm{SI}}[t]$. Accordingly, the SINR constraint, $\gamma_k[t]\ge\gamma_{\rm th}$, is equivalent to
\begin{align}\label{eq:SINR_DC_form}
\tilde{\mathbf{w}}_k^{\mathsf H}[t]{\mathbf{D}_k\mathbf{H}_k[t]\mathbf{D}_k}\tilde{\mathbf{w}}_k[t]
-\gamma_{\rm th}
&\sum_{j\neq k}
\tilde{\mathbf{w}}_j^{\mathsf H}[t]\mathbf{D}_j\mathbf{H}_k[t]\mathbf{D}_j\tilde{\mathbf{w}}_j[t]\notag\\
& \quad \ge\gamma_{\rm th}\left({I_k^{\mathrm{SI}}[t]}+\sigma^2\right).
\end{align}
This motivates the use of SDR for beamforming updates and the use of SCA for association updates.

\subsubsection{Geometric and LoS-Induced Non-Convexity}

For the  $s$-th ST, the CRB admits the Schur-complement form in \eqref{eq:crb} and depends on the unified cross-field array response in \eqref{eq:unified_response}, which involves nonlinear functions of the propagation distance and the corresponding array manifold. Moreover, the CRB requires matrix inversion of strongly coupled FIM blocks derived in Appendix \ref{App:Derivation_CRB_Matrix}. These geometric nonlinearities render $\mathrm{CRB}_s$ intrinsically non-convex. Furthermore, blockage and LoS conditions determine the participating AP set via $\mathcal{M}_k=\{m\in\mathcal{M}:p_{\rm LoS}(r_{m,k})\ge p_{\rm th}\}$, which induces a geometry-dependent serving topology. Since the LoS probability is distance-dependent, both the effective channel structure and sensing FIM become state-dependent. Consequently, cross-field propagation and LoS-constrained participation jointly introduce tightly coupled geometric-topological non-convexity.

\subsection{Problem Transformation and Reformulation}\label{subsec:problem_reformulation}

In this subsection, we first relax the binary association coefficients and then derive a set of tractable reformulations and convex surrogates in \eqref{prob:P0}.

\subsubsection{Integer Relaxation for Association Coefficients}

We note that the association coefficients satisfy $\delta_{m,k}\in\{0,1\}$, which induces the combinatorial non-convexity of \eqref{prob:P0}. Thus, in the first step, we relax them to continuous variables $ 0 \le \delta_{m,k} \le 1$. Under this relaxation, $\delta_{m,k}$ acts as a soft participation weight. This yields a continuous but still non-convex problem, due to the coupling between $\delta_{m,k}$ and $\mathbf{w}_{m,k}$ in the SINR and CRB expressions.

\subsubsection{SCA Reformulation for SINR Constraints}
For each UE, the SINR constraint in \eqref{prob:P0b} can be written as
\begin{align}\label{eq:SINR_rewrite}
\left|\sum_{m=1}^{M} \delta_{m,k}\,\mathbf{h}_{m,k}^{\mathsf H}[t] \mathbf{w}_{m,k}[t]\right|^2
\geq\gamma_{\rm th}\left(I_k[t] + \sigma^2\right).
\end{align}
For fixed relaxed association coefficients $\delta_{m,k}$ and fixed sensing waveforms, using the stacked representation introduced in the problem analysis, \eqref{eq:SINR_rewrite} admits a compact form as
\begin{align}\label{eq:SINR_DC_compact}
f_k(\mathbf{w};\boldsymbol{\delta})-g_k(\mathbf{w};\boldsymbol{\delta})
\ge \gamma_{\rm th}\!\left(I_k^{\mathrm{SI}}[t]+\sigma^2\right),
\end{align}
where $f_k(\mathbf{w};\boldsymbol{\delta})$ corresponds to the desired signal power and $g_k(\mathbf{w};\boldsymbol{\delta})$ collects the scaled multiuser interference term. After SDR lifting with fixed $\boldsymbol{\delta}$, both $f_k$ and $g_k$ become affine in the covariance variables, whereas for fixed beamforming covariances the dependence on $\boldsymbol{\delta}$ remains non-convex and is handled by SCA.

\subsubsection{SDR Relaxation for Beamforming Covariances}

To facilitate a unified treatment of the SINR and CRB constraints, we adopt a lifted covariance representation based on stacked beamformers. Specifically, we define the stacked beamformer, $\tilde{\mathbf{w}}_k[t]\in\mathbb{C}^{N_{\rm tot}\times1}$, as in \eqref{eq:signal_quadratic_form} and introduce the lifted covariance matrix, $\mathbf{W}_k[t]\triangleq \tilde{\mathbf{w}}_k[t]\tilde{\mathbf{w}}_k^{\mathsf H}[t]\succeq\mathbf{0}$. The rank-one constraint, $\mathrm{rank}(\mathbf{W}_k[t])=1$, enforces a single-stream beamformer and renders the problem non-convex. By temporarily relaxing this constraint, we obtain an SDR of the original formulation \cite{5447068}. If the resulting solution is rank one, the optimal stacked beamformer can be directly recovered via eigendecomposition; otherwise, a feasible rank-one approximation is constructed using Gaussian randomization. Under the lifted representation, the per-AP power constraint can be equivalently expressed using a block-selection matrix,
$\mathbf{E}_m\triangleq \mathrm{blkdiag}(\mathbf{0},\ldots,\mathbf{I}_{N_m},\ldots,\mathbf{0})$,
which extracts the $m$-th AP block as
\begin{align}\label{eq:power_constraint_W}
\sum_{k\in\mathcal{K}} \operatorname{tr}\!\left(\mathbf{E}_m \mathbf{W}_k[t]\right)
\le P_{\mathrm{max}},\quad \forall m.
\end{align}

\subsubsection{Convex Surrogates for CRB Constraints}\label{subsubsec:crb_convex_surrogates}

For sensing performance, the CRB requirement $\mathrm{CRB}_{s}(\{\mathbf{W}_k\})\leq \epsilon_{\mathrm{th}}$ can be equivalently expressed as
\begin{align}\label{eq:CRB_exact}
\mathbf{J}_{11,s}-\mathbf{J}_{12,s}\mathbf{J}_{22,s}^{-1}\mathbf{J}_{12,s}^{\mathsf H}\succeq
\epsilon_{\mathrm{th}}^{-1}\mathbf{I},\quad \forall s\in\mathcal{S}.
\end{align}
This constraint is non-convex due to the matrix inverse and the bilinear coupling between FIM sub-blocks.
To obtain a tractable formulation, we adopt a convex surrogate by retaining only the dominant FIM block and enforcing
\begin{align}\label{eq:CRB_surrogate}
\mathbf{J}_{11,s}(\mathbf{W})
\succeq
\epsilon_{\mathrm{th}}^{-1}\mathbf{I},\quad \forall s\in\mathcal{S},
\end{align}
where $\mathbf{W}\triangleq\{\mathbf{W}_k\}_{k\in\mathcal{K}}$ denotes the stacked lifted covariances. This surrogate yields a convex feasible set under the lifted beamforming representation and serves as a tractable proxy for the exact CRB constraint. Specifically, since $\mathbf{J}_{11,s}-\mathbf{J}_{12,s}\mathbf{J}_{22,s}^{-1}\mathbf{J}_{12,s}^{\mathsf H}\preceq \mathbf{J}_{11,s}$ in the Loewner order, \eqref{eq:CRB_surrogate} does not generally imply \eqref{eq:CRB_exact} and should be interpreted as a tractable surrogate. The approximation becomes tighter when the coupling block $\mathbf{J}_{12,s}$ is weak. To keep the sensing objective explicit, we adopt
\begin{align}\label{eq:phi_def}
\phi_s\!\big(\mathbf{J}_{11,s}(\mathbf{W})\big)
\triangleq
-\log\det\!\big(\mathbf{J}_{11,s}(\mathbf{W})+\epsilon_{\phi}\mathbf{I}\big),
\end{align}
where $\epsilon_{\phi}>0$ ensures numerical stability.

\subsection{Optimization Based Solution to the Transformed ISAC Problem}\label{subsec:opt_solution}

With the aforementioned integer relaxation, we develop the B2S algorithm to solve the resulting ISAC design problem. Since the relaxed association coefficients and beamforming variables remain coupled \cite{5756489}, B2S adopts a BCD framework that alternates between an SDR based beamforming covariance update and an SCA based association-coefficient update.

\subsubsection{B2S Decomposition and Convex Surrogates}

During the $i$-th B2S iteration, we alternate between two blocks:
\begin{itemize}
\item \textit{Beamforming-covariance update:} Optimize the lifted covariances with fixed $\boldsymbol{\delta}=\boldsymbol{\delta}^{(i)}$;
\item \textit{Association-coefficient update:} Optimize $\boldsymbol{\delta}$ with fixed beamforming covariances.
\end{itemize}
Using the stacked beamformer $\tilde{\mathbf{w}}_k[t]\in\mathbb{C}^{N_{\rm tot}\times 1}$, we adopt the lifted covariance matrix $\mathbf{W}_k[t]$. 
For fixed $\boldsymbol{\delta}^{(i)}$, let $\mathbf{D}_k^{(i)}$ be constructed from $\boldsymbol{\delta}^{(i)}$ as in \eqref{eq:SINR_DC_form}, and let $\mathbf{H}_k[t]$ denote the stacked effective channel matrix. Then, the desired signal power and multiuser interference terms at the $k$-th UE are written as
\begin{align}\label{eq:Sk_trace}
S_k^{(i)}[t]=\operatorname{tr}\left(
\mathbf{D}_k^{(i)}\mathbf{H}_k[t]\mathbf{D}_k^{(i)}\,\mathbf{W}_k[t]
\right)\end{align}
and
\begin{align}\label{eq:Ik_trace}
I^{\mathrm{MU},(i)}_k[t]=
\sum_{j\neq k}\operatorname{tr}\left(\mathbf{D}_j^{(i)}\mathbf{H}_k[t]\mathbf{D}_j^{(i)}\,\mathbf{W}_j[t]
\right),
\end{align}
respectively. Hence, following \eqref{eq:SINR_DC_form}, the SINR constraint $\gamma_k[t]\geq\gamma_{\mathrm{th}}$ can be equivalently reformulated as
\begin{align}\label{eq:SINR_DC_compact_1}
f_k(\mathbf{W};\boldsymbol{\delta}^{(i)})-g_k(\mathbf{W};\boldsymbol{\delta}^{(i)})
\geq\gamma_{\rm th}\!\left(I_k^{\mathrm{SI}}[t]+\sigma^2\right),
\end{align}
where $f_k(\mathbf{W};\boldsymbol{\delta}^{(i)})\triangleq S_k^{(i)}[t]$ denotes the desired signal power and $g_k(\mathbf{W};\boldsymbol{\delta}^{(i)})\triangleq \gamma_{\rm th} I_k^{\mathrm{MU},(i)}[t]$ denotes the scaled multiuser interference term. For fixed $\boldsymbol{\delta}^{(i)}$, both terms are affine in $\mathbf{W}$.

\subsubsection{B2S Beamforming-Covariance Update via SDR}

For sensing performance, we note that the exact CRB constraint \eqref{eq:CRB_exact} is non-convex. Following Section \ref{subsec:problem_reformulation}, we adopt the convex surrogate $\mathbf{J}_{11,s}(\mathbf{W},\boldsymbol{\delta}^{(i)}) \succeq \epsilon_{\mathrm{th}}^{-1}\mathbf{I},\ \forall s \in \mathcal{S}$, as it provides a tractable sensing proxy for the exact CRB requirement. Here, $\mathbf{J}_{11,s}(\mathbf{W},\boldsymbol{\delta}^{(i)})$ is affine in the lifted beamforming covariances for fixed $\boldsymbol{\delta}^{(i)}$. Combining these ingredients, the $i$-th SDR based beamforming subproblem is formulated as
\begin{subequations}\label{prob:W_subproblem}
\begin{align}
\mathcal{P}_{\mathbf{W}}^{(i)}:\quad
&\min_{\{\mathbf{W}_{k}\succeq\mathbf{0}\}}
\quad\sum_{s=1}^{S}\phi_s\left(\mathbf{J}_{11,s}(\mathbf{W},\boldsymbol{\delta}^{(i)})\right) \label{eq:obj_phi}\\
\text{s.t.}\quad
& f_k(\mathbf{W};\boldsymbol{\delta}^{(i)}) - g_k(\mathbf{W};\boldsymbol{\delta}^{(i)}) \nonumber\\
& \qquad \qquad \qquad \geq \gamma_{\mathrm{th}}\!\left(I^{\mathrm{SI}}_k + \sigma^2\right),~\forall k, \label{eq:sca_sinr}\\
& \sum_{k\in\mathcal{K}} \operatorname{tr}\!\left(\mathbf{E}_m\mathbf{W}_{k}\right)\le P_{\max},~\forall m,\label{eq:sca_power}\\
&  \mathbf{J}_{11,s}(\mathbf{W},\boldsymbol{\delta}^{(i)})\succeq \epsilon_{\mathrm{th}}^{-1}\mathbf{I},~\forall s.
\label{eq:sca_crb}
\end{align}
\end{subequations}
We note that problem \eqref{prob:W_subproblem} is a semidefinite program (SDP) that can be efficiently solved by off-the-shelf solvers such as CVX. Finally, the rank-one constraints, i.e., $\mathrm{rank}(\mathbf{W}_k)=1$, are temporarily dropped to yield the SDR, and the corresponding beamformers are recovered in a post-processing step.

\subsubsection{B2S Association-Coefficient Update via SCA}

With fixed beamforming covariances from \eqref{prob:W_subproblem}, the optimization over $\boldsymbol{\delta}$ remains non-convex due to its quadratic coupling in the SINR expressions. For the $k$-th UE, the SINR constraint can be rewritten in DC form as $u_k(\boldsymbol{\delta})-v_k(\boldsymbol{\delta}) \ge 0$, where both $u_k(\boldsymbol{\delta})$ and $v_k(\boldsymbol{\delta})$ are convex quadratic functions of $\boldsymbol{\delta}$. To obtain a convex inner approximation, we linearize the convex term $u_k(\boldsymbol{\delta})$ at the current iterate $\boldsymbol{\delta}^{(i)}$ as
\begin{align}\label{eq:uk_taylor_delta}
u_k(\boldsymbol{\delta})
\!\geq\!\hat{u}_k(\boldsymbol{\delta}\!\mid\! \boldsymbol{\delta}^{(i)})
\!\triangleq\! u_k(\boldsymbol{\delta}^{(i)})
\!+\!\nabla u_k(\boldsymbol{\delta}^{(i)})^{\mathsf T}
\!\left(\boldsymbol{\delta}\!-\!\boldsymbol{\delta}^{(i)}\right).
\end{align}
This yields the convex inner approximation as 
\begin{align}\label{eq:SINR_SCA_delta}
v_k(\boldsymbol{\delta})-\hat{u}_k(\boldsymbol{\delta}\mid \boldsymbol{\delta}^{(i)})\leq0,\quad \forall k.
\end{align}
To stabilize the BCD iterations, we optionally include a proximal regularization term in the association-coefficient update. Moreover, to maintain consistency with the sensing-aware transformed problem, we incorporate sensing-side surrogate constraints into the $\boldsymbol{\delta}$-update with fixed beamforming covariances ${\mathbf W}^{(i+1)}$. The resulting SCA subproblem is given by
\begin{subequations}\label{prob:delta_sca_subproblem}
\begin{align}
\mathcal{P}_{\delta}^{(i)}:\quad
&\min_{\boldsymbol{\delta}}
\widehat{\Phi}_{\mathrm{sen}}\left(\boldsymbol{\delta}\mid \boldsymbol{\delta}^{(i)}\right)
+\Psi(\boldsymbol{\delta})
+\frac{\tau_\delta}{2}\|\boldsymbol{\delta}-\boldsymbol{\delta}^{(i)}\|^2
\label{prob:delta_sca_subproblem_obj}
\\
\mathrm{s.t.}\quad
&
v_k(\boldsymbol{\delta})
-
\hat{u}_k(\boldsymbol{\delta}\mid \boldsymbol{\delta}^{(i)})
\le 0,\quad \forall k,
\label{prob:delta_sca_subproblem_sinr}
\\
&
\epsilon_{\mathrm{th}}^{-1}\mathbf{I}
-\widehat{\mathbf{J}}_{11,s}\!\left(\boldsymbol{\delta}\mid\boldsymbol{\delta}^{(i)}\right)
\preceq \mathbf{0},\quad \forall s,
\label{prob:delta_sca_subproblem_crb}
\\
&
0\le \delta_{m,k}\le \xi_{m,k},\quad \forall m,k,
\label{prob:delta_sca_subproblem_box}
\end{align}
\end{subequations}
where $\Psi(\boldsymbol{\delta})$ denotes any additional convex regularizer on the association coefficients and $\tau_\delta\ge 0$ is a proximal parameter. Here,
$\widehat{\mathbf{J}}_{11,s}(\boldsymbol{\delta}\mid\boldsymbol{\delta}^{(i)})$ denotes a convex surrogate of $\mathbf{J}_{11,s}({\mathbf W}^{(i+1)},\boldsymbol{\delta})$, i.e.,
$\widehat{\mathbf{J}}_{11,s}(\boldsymbol{\delta}\mid\boldsymbol{\delta}^{(i)})
\preceq
\mathbf{J}_{11,s}({\mathbf W}^{(i+1)},\boldsymbol{\delta})$,
which yields a conservative semidefinite sensing constraint in the SCA update. We clarify that problem \eqref{prob:delta_sca_subproblem} is a convex program and can be solved efficiently by standard solvers.

\subsubsection{Overall B2S Procedure and Recovery}

Starting from a feasible initialization $\big({\mathbf{W}}^{(0)},\boldsymbol{\delta}^{(0)}\big)$, the proposed B2S algorithm alternates between solving \eqref{prob:W_subproblem} and \eqref{prob:delta_sca_subproblem}. Under standard SCA conditions, the adopted surrogate objective is non-increasing over iterations, and the sequence generated by B2S converges to a stationary point of the relaxed transformed problem. After convergence, the beamforming vectors are recovered from the SDR solution via eigendecomposition, and the relaxed association coefficients are mapped to binary values via thresholding or feasibility refinement to obtain a feasible design for the original MINLP problem. Finally, we verify \eqref{eq:CRB_exact} \textit{a posteriori} for the recovered beamformers. For clarity, the overall procedure is summarized in \textbf{Algorithm \ref{alg:BCD_SCA_SDR}}.

\begin{algorithm}[!t]
\begin{small}
\caption{B2S Algorithm for Transformed ISAC Problem}
\label{alg:BCD_SCA_SDR}
\begin{algorithmic}[1]
\State \textbf{Input:} CSI, visibility information, system parameters, thresholds $\gamma_{\rm th}$, $\epsilon_{\rm th}$, tolerance $\epsilon_{\mathrm{BCD}}$, maximum iterations $I_{\max}$.
\State \textbf{Initialize:} Construct relaxed association coefficients $\boldsymbol{\delta}^{(0)}\!\in[0,1]$; generate feasible stacked covariances $\mathbf{W}^{(0)}=\{\mathbf{W}_k^{(0)}\succeq\mathbf{0}\}_{k\in\mathcal{K}}$; set BCD iteration index $i\gets 0$.
\State Evaluate the initial objective $\mathcal{J}^{(0)}$.

\For{$i=0,1,\ldots,I_{\max}-1$}

\Comment{Beamforming-covariance update (SDR step)}
    \State Construct effective matrices $\mathbf{H}_k[t]$ and selection matrices $\mathbf{D}_k^{(i)}$ from $\boldsymbol{\delta}^{(i)}$.
    \State Form the beamforming SDP subproblem $\mathcal{P}_{\mathbf{W}}^{(i)}$ with fixed $\boldsymbol{\delta}^{(i)}$.
    \State Solve $\mathcal{P}_{\mathbf{W}}^{(i)}$ to obtain $\mathbf{W}^{(i+1)}=\{\mathbf{W}_k^{(i+1)}\}_{k\in\mathcal{K}}$.

\Comment{Association-coefficient update (SCA step)}
    \State Build DC components $u_k(\boldsymbol{\delta})$ and $v_k(\boldsymbol{\delta})$ for the SINR constraints using fixed $\mathbf{W}^{(i+1)}$.
    \State Compute affine inner approximations $\hat{u}_k(\boldsymbol{\delta}\mid\boldsymbol{\delta}^{(i)})$.
    \State Form the convex SCA subproblem $\mathcal{P}_{\delta}^{(i)}$.
    \State Solve $\mathcal{P}_{\delta}^{(i)}$ to obtain $\boldsymbol{\delta}^{(i+1)}$.

\Comment{Stopping criterion}
    \State Evaluate objective $\mathcal{J}^{(i+1)}$.
    \If{$\dfrac{|\mathcal{J}^{(i+1)}-\mathcal{J}^{(i)}|}{\max\{1,|\mathcal{J}^{(i)}|\}} \le \epsilon_{\mathrm{BCD}}$}
        \State \textbf{break}
    \EndIf

\EndFor

\Comment{Rank-one recovery from SDR solution}
\If{all $\mathbf{W}_k^\star$ are rank-one}
    \State Recover stacked beamformers $\{\tilde{\mathbf{w}}_k^\star\}$ from $\{\mathbf{W}_k^\star\}$ via eigendecomposition and extract per-AP blocks $\{\mathbf{w}_{m,k}^\star\}$.
\Else
    \State Apply Gaussian randomization $\mathbf{v}_k^{(n)}\sim\mathcal{CN}(\mathbf{0},\mathbf{W}_k^\star)$, normalize candidates to satisfy per-AP power constraints, and select the best feasible candidate.
\EndIf

\Comment{Binary mapping of relaxed association coefficients}
\State Map $\boldsymbol{\delta}^{\star}$ to binary association coefficients.
\State Verify \eqref{eq:CRB_exact} for the recovered solution.

\State \textbf{Output:} Final beamformers $\{\mathbf{w}_{m,k}^{\star}\}$ and binary association coefficients $\{\delta_{m,k}^{\star}\}$.
\end{algorithmic}
\end{small}
\end{algorithm}

\section{Graph Transformer Network for Structured ISAC Representation}\label{sec:GTN_main}

Building upon the cross-field THz CF-ISAC model in Section \ref{sec:II_system} and the optimization reformulations in Section \ref{sec:opt_solution}, we redesign the GTN as an \emph{optimization-aware structured representation module}. Instead of directly generating beamforming decisions \cite{10419173}, the GTN learns permutation-equivariant embeddings that encapsulate cross-field wavefront geometry, near-/far-field coupling, blockage visibility, and sensing relevance, which interface with the B2S solver to enable warm-start initialization and feasibility-aware weighting.

\subsection{Heterogeneous Interaction Graph and Features}\label{subsec:GTN_graph}

The CF-ISAC system involves strongly coupled many-to-many interactions over heterogeneous cross-field THz channels. To capture such interactions under dynamic system topologies, we employ a permutation-equivariant GTN that models cross-node dependencies via attention and produces geometry- and channel-aware embeddings. We construct a heterogeneous interaction graph $\mathcal{G}=(\mathcal{V},\mathcal{E})$ with $|\mathcal{V}|=MK$ nodes, where node $i=\mathcal{I}(m,k)\in\{1,\ldots,MK\}$ corresponds to AP-UE pair $(m,k)$.

\subsubsection{Physics-Aware Node Features}

For node $i=\mathcal{I}(m,k)$, we define a physics-aware feature vector $\bm{\chi}_{m,k}\in\mathbb{R}^{d_0}$ that encapsulates the cross-field propagation geometry, LoS visibility, THz-band pathloss and molecular absorption, as well as the cross-field array response derived in Section \ref{sec:II_system}, given by
\begin{align}\label{eq:GTN_chi_expand}
&\bm{\chi}_{m,k}\notag\\
&\triangleq
\Big[
\underbrace{\tfrac{r_{m,k}}{R_m}}_{\text{cross-field ratio}}, \underbrace{\mathbb{I}\!\left\{r_{m,k}<R_m\right\}}_{\text{near-field flag}}, \underbrace{p_{\mathrm{LoS}}(r_{m,k})}_{\text{visibility}}, \underbrace{\log L_{m,k}}_{\text{pathloss+absorption}}, \notag\\
&\underbrace{\operatorname{Re}\{\tilde{\beta}_{m,k}\},\operatorname{Im}\{\tilde{\beta}_{m,k}\}}_{\text{effective complex gain}}, \underbrace{\kappa_{m,k}}_{\text{phase-curvature index}}, \underbrace{\zeta^{(i)}_{m,k}}_{\text{B2S sensing cue}} \Big]^{\mathsf T},
\end{align}
where $\mathbb{I}\{\cdot\}$ is the indicator function, $\tilde{\beta}_{m,k}\triangleq \beta_{m,k}e^{-j\tfrac{2\pi f}{c}r_{m,k}}$ and \begin{align}\label{eq:GTN_curvature_index}
\kappa_{m,k}
\triangleq
\frac{1}{N_m}\left\|
\mathbf{a}(\theta_{m,k},r_{m,k})-e^{-j\tfrac{2\pi f}{c}r_{m,k}}\mathbf{1}_{N_m}
\right\|^2.
\end{align}
Here, $\mathbf{1}_{N_m}\in\mathbb{R}^{N_m}$ denotes the all-ones vector. For convenience, we define the shorthand $\mathbf{a}_{m,k}\triangleq \mathbf{a}(\theta_{m,k},r_{m,k})$. During the $i$-th B2S iteration in Section \ref{sec:opt_solution}, the beamforming covariances $\mathbf{W}^{(i)}=\{\mathbf{W}_k^{(i)}\}_{k\in\mathcal K}$ and relaxed associations $\boldsymbol{\delta}^{(i)}$ induce the per-AP covariance, given by
\begin{align}\label{eq:GTN_Qm_BCD}
\mathbf{Q}_m^{(i)}
\triangleq
\sum_{k=1}^{K}\left(\delta_{m,k}^{(i)}\right)^{2}\,
\mathbf{S}_m\mathbf{W}_k^{(i)}\mathbf{S}_m^{\mathsf H}
+\mathbf{Q}^{\mathrm{sen}}_m,
\end{align}
where $\mathbf{S}_m\in\{0,1\}^{N_m\times N_{\rm tot}}$ extracts the $m$-th AP block and $\mathbf{Q}^{\mathrm{sen}}_m$ is the covariance induced by the dedicated sensing pilots. Consistent with the surrogate sensing formulation in Section \ref{sec:opt_solution}, the FIM block $\mathbf{J}_{11,m,s}$ is affine in $\mathbf{Q}_m^{(i)}$.

To quantify the sensing relevance of node $(m,k)$, we use a marginal-information proxy obtained by removing the $k$-th UE covariance contribution at AP $m$, i.e.,
$\mathbf{Q}_{m\setminus k}^{(i)}\triangleq
\mathbf{Q}_m^{(i)}-\left(\delta_{m,k}^{(i)}\right)^{2}
\mathbf{S}_m\mathbf{W}_k^{(i)}\mathbf{S}_m^{\mathsf H}$,
and define the B2S-consistent sensing cue as
\begin{align}\label{eq:GTN_sensing_cue}
\zeta_{m,k}^{(i)}
\triangleq
\sum_{s=1}^{S}\delta_{m,s}\,
\Big[
&{\log\det\big(\mathbf{J}_{11,m,s}(\mathbf{Q}_m^{(i)})+\epsilon_0\mathbf{I}\big)}\notag\\
&-\log\det\big(\mathbf{J}_{11,m,s}(\mathbf{Q}_{m\setminus k}^{(i)})+\epsilon_0\mathbf{I}\big)
\Big],
\end{align}
where $\epsilon_0>0$ ensures numerical stability. This choice is aligned with the surrogate sensing objective in \eqref{eq:phi_def}. At initialization, we set $i=0$ and compute $\zeta^{(0)}_{m,k}$ using the GTN warm-start covariances that will be defined in \eqref{eq:GTN_warm_start}. Therefore, we initialize the node's hidden state as
\begin{align}\label{eq:GTN_h0}
\mathbf{h}_i^{(0)} \triangleq \mathbf{h}^{(0)}_{m,k}
=\phi_{\mathrm{in}}(\bm{\chi}_{m,k})
=\mathrm{LN}\!\big(\mathbf{W}_{\mathrm{in}}\bm{\chi}_{m,k}+\mathbf{b}_{\mathrm{in}}\big),
\end{align}
where $\phi_{\mathrm{in}}(\cdot)$ denotes a learnable input embedding function, $\mathbf{W}_{\mathrm{in}}\in\mathbb{R}^{d_h\times d_0}$ and $\mathbf{b}_{\mathrm{in}}\in\mathbb{R}^{d_h}$ are trainable parameters,
$d_h$ denotes the hidden feature dimension of the GTN, and $\mathrm{LN}(\cdot)$ denotes layer normalization.

\subsubsection{Heterogeneous Edge Sets and Edge Features}

We note that two nodes are connected if they share the same AP or the same UE. This yields two heterogeneous edge types, given by
\begin{equation}\label{eq:GTN_edges_AP}
\mathcal{E}^{\mathrm{AP}}
=
\big\{
(\mathcal{I}(m,k),\mathcal{I}(m,k'))~\big|~k\neq k'
\big\}
\end{equation}
and
\begin{equation}\label{eq:GTN_edges_UE}
\mathcal{E}^{\mathrm{UE}}
=
\big\{
(\mathcal{I}(m,k),\mathcal{I}(m',k))~\big|~m\neq m'
\big\},
\end{equation}
where $\mathcal{E}=\mathcal{E}^{\mathrm{AP}}\cup\mathcal{E}^{\mathrm{UE}}$. Here, we define and use the normalized array-similarity score as $\varsigma(\mathbf{x},\mathbf{y})\triangleq \frac{|\mathbf{x}^{\mathsf H}\mathbf{y}|}{\|\mathbf{x}\|\|\mathbf{y}\|}$, as well as defining the normalized relative direction as $\hat{\mathbf{q}}_{m,k}\triangleq\frac{\mathbf{u}_k-\mathbf{p}_m}{\|\mathbf{u}_k-\mathbf{p}_m\|}$.

For an AP-type edge corresponding to the AP-UE pairs $i=\mathcal{I}(m,k)$ and $j=\mathcal{I}(m,k')$, we define
\begin{align}\label{eq:GTN_edge_AP_feat}
\mathbf{e}^{\mathrm{AP}}_{ij}
\triangleq
\Big[
&\Delta\hat{\mathbf{q}}_{m,kk'}^{\mathsf T},~
\Delta \tfrac{r_{m,k}}{R_m},~
\Delta p_{\mathrm{LoS}},~
\varsigma(\mathbf{a}_{m,k},\mathbf{a}_{m,k'})
\Big]^{\mathsf T},
\end{align}
where $\Delta\hat{\mathbf{q}}_{m,kk'}\triangleq\hat{\mathbf{q}}_{m,k}-\hat{\mathbf{q}}_{m,k'}$,
$\Delta \tfrac{r_{m,k}}{R_m}\triangleq\tfrac{r_{m,k}}{R_m}-\tfrac{r_{m,k'}}{R_m}$,
and $\Delta p_{\mathrm{LoS}}\triangleq p_{\mathrm{LoS}}(r_{m,k})-p_{\mathrm{LoS}}(r_{m,k'})$. Similarly, for a UE-type edge corresponding to the AP-UE pairs $i=\mathcal{I}(m,k)$ and $j=\mathcal{I}(m',k)$, we define
\begin{align}\label{eq:GTN_edge_UE_feat}
\mathbf{e}^{\mathrm{UE}}_{ij}
\!\triangleq\!
\Big[
&\Delta\hat{\mathbf{q}}_{mm',k}^{{\mathsf T}},~
\tfrac{r_{m,k}}{R_m},~
\tfrac{r_{m',k}}{R_{m'}},~
p_{\mathrm{LoS}}(r_{m,k}), \notag\\
& \qquad\qquad\qquad p_{\mathrm{LoS}}(r_{m',k}),~
\varsigma(\mathbf{a}_{m,k},\mathbf{a}_{m',k})
\Big]^{\mathsf T},
\end{align}
where $\Delta\hat{\mathbf{q}}_{mm',k}\triangleq\hat{\mathbf{q}}_{m,k}-\hat{\mathbf{q}}_{m',k}$.

\subsection{Heterogeneous Graph Transformer Encoder}\label{subsec:GTN_arch}

We adopt an $L$-layer heterogeneous graph transformer \cite{10.1145/3366423.3380027} with $C$ attention heads at each AP-/UE-type edge.

\subsubsection{Type-Specific Multi-Head Attention with Geometry Bias}

From $\mathbf{h}_i^{(0)}$ defined in \eqref{eq:GTN_h0}, the node representation $\mathbf{h}_i^{(l)}$ is updated across the $L$ GTN layers. For each layer $l\in\{0,\ldots,L-1\}$, attention head $c\in\{1,\ldots,C\}$, and directed neighbor $j\in\mathcal{N}^{\circ}(i)$, $\circ\in\{\mathrm{AP},\mathrm{UE}\}$, we obtain
\begin{align}\label{eq:GTN_qkv}
\mathbf{q}^{\circ,(l)}_{c,i}&\!=\!\mathbf{W}^{\circ,Q}_{c}\mathbf{h}^{(l)}_{i}\!,\!\mathbf{k}^{\circ,(l)}_{c,j}\!=\!\mathbf{W}^{\circ,K}_{c}\mathbf{h}^{(l)}_{j}\!,\!\mathbf{v}^{\circ,(l)}_{c,j}
\!=\!\mathbf{W}^{\circ,V}_{c}\mathbf{h}^{(l)}_{j},
\end{align}
where $\mathbf{W}^{\circ,Q}_{c}\in\mathbb{R}^{d_a \times d_h}$, $\mathbf{W}^{\circ,K}_{c}\in\mathbb{R}^{d_a \times d_h}$, and $\mathbf{W}^{\circ,V}_{c}\in\mathbb{R}^{d_a \times d_h}$ are learnable projection matrices associated with the $\circ$-type edge and attention head $c$, and $d_a$ denotes the per-head query/key/value embedding dimension. To inject physics-aware relative information, we define a learned geometry bias via a multi-layer perceptron
$\psi_{\circ}(\cdot)$ as
\begin{align}\label{eq:GTN_bias}
b^{\circ,(l)}_{c,ij}=\mathbf{u}^{\circ,(l){\mathsf T}}_{c}\,
\psi_{\circ}\big(\mathbf{e}^{\circ}_{ij}\big),~~
\psi_{\circ}(\mathbf{x}) =
\rho(\mathbf{A}_{\circ}\mathbf{x}+\mathbf{a}_{\circ}),
\end{align}
where $\mathbf{A}_{\circ}\in\mathbb{R}^{d_b\times d_e}$ and $\mathbf{a}_{\circ}\in\mathbb{R}^{d_b}$ are learnable parameters of the edge encoder, $\mathbf{u}^{\circ,(l)}_{c}\in\mathbb{R}^{d_b}$ is a head- and layer-specific projection vector, and $\rho(\cdot)$ is an elementwise nonlinearity. The attention logit and coefficient are given by
\begin{equation}\label{eq:GTN_logit}
s^{\circ,(l)}_{c,ij}=\frac{1}{\sqrt{d_a}}\mathbf{q}^{\circ,(l){\mathsf T}}_{c,i}\mathbf{k}^{\circ,(l)}_{c,j}+b^{\circ,(l)}_{c,ij}
\end{equation}
and
\begin{equation}\label{eq:GTN_alpha}
\alpha^{\circ,(l)}_{c,ij}=\frac{e^{s^{\circ,(l)}_{c,ij}}}{\sum_{n\in\mathcal{N}^{\circ}(i)}e^{s^{\circ,(l)}_{c,in}}}.
\end{equation}

\subsubsection{Type-Wise Message Aggregation and Gated Fusion}

The type-wise aggregated message for node $i$ is given by
\begin{align}\label{eq:GTN_message}
\mathbf{m}^{\circ,(l)}_{c,i}
=\sum_{j\in\mathcal{N}^{\circ}(i)}
\alpha^{\circ,(l)}_{c,ij}\,\mathbf{v}^{\circ,(l)}_{c,j},\mathbf{m}^{\circ,(l)}_{i}
=\bigoplus_{c=1}^{C}\mathbf{m}^{\circ,(l)}_{c,i},
\end{align}
where $\bigoplus$ denotes concatenation across heads. To reflect the distinct roles of AP-type and UE-type edges, we adopt a learnable gate to fuse two messages, expressed as
\begin{equation}
\boldsymbol{\eta}^{(l)}_{i}
=
\sigma\!\left(
\mathbf{W}^{(l)}_{\eta}
\big[
\mathbf{m}^{\mathrm{AP},(l)}_{i}\|
\mathbf{m}^{\mathrm{UE},(l)}_{i}\|
\mathbf{h}^{(l)}_{i}
\big]
+\mathbf{b}^{(l)}_{\eta}
\right)
\label{eq:GTN_gate}
\end{equation}
and
\begin{equation} \label{eq:GTN_fuse}
\mathbf{m}^{(l)}_{i}
=
\boldsymbol{\eta}^{(l)}_{i}\odot\mathbf{m}^{\mathrm{AP},(l)}_{i}
+
\big(\mathbf{1}-\boldsymbol{\eta}^{(l)}_{i}\big)\odot\mathbf{m}^{\mathrm{UE},(l)}_{i},
\end{equation}
respectively, where $\sigma(\cdot)$ denotes the sigmoid function and $\odot$ denotes the Hadamard product.

\subsubsection{Transformer Block Update}

We update node embeddings with a standard transformer block as
\begin{subequations}\label{eq:GTN_transformer}
\begin{align}
\tilde{\mathbf{h}}^{(l+1)}_{i}
&=
\mathrm{LN}\left(
\mathbf{h}^{(l)}_{i}
+
\mathbf{W}^{(l)}_{o}\mathbf{m}^{(l)}_{i}
\right),
\label{eq:GTN_res1}\\
\mathbf{h}^{(l+1)}_{i}
&=
\mathrm{LN}\left(
\tilde{\mathbf{h}}^{(l+1)}_{i}
+
\mathrm{FFN}^{(l)}\left(\tilde{\mathbf{h}}^{(l+1)}_{i}\right)
\right),
\label{eq:GTN_res2}\\
\mathrm{FFN}^{(l)}(\mathbf{x})
&=
\mathbf{W}^{(l)}_{2}\,\rho\left(\mathbf{W}^{(l)}_{1}\mathbf{x}+\mathbf{b}^{(l)}_{1}\right)
+\mathbf{b}^{(l)}_{2}.
\label{eq:GTN_ffn}
\end{align}
\end{subequations}
All operations depend on $\mathcal{G}$ only and are permutation equivariant to AP/UE indexing.

\subsection{Permutation-Invariant Readout}\label{subsec:GTN_readout}

After $L$ layers, the embeddings $\mathbf{h}^{(L)}_{m,k}$ are aggregated via attention pooling to form AP-level and UE-level representations aligned with the B2S modules. For each AP, we pool all nodes $\{(m,k)\}_{k=1}^{K}$ as
\begin{align}\label{eq:GTN_ap_pool}
\Omega_{m,k}=\frac{e^{\mathbf{a}^{\mathsf T}\mathbf{h}^{(L)}_{m,k}}}{\sum_{k'=1}^{K}e^{\mathbf{a}^{\mathsf T}\mathbf{h}^{(L)}_{m,k'}}},
~~\mathbf{g}^{\mathrm{AP}}_{m}=\sum_{k=1}^{K}\Omega_{m,k}\mathbf{h}^{(L)}_{m,k}.
\end{align}
For each UE, we pool all nodes $\{(m,k)\}_{m=1}^{M}$ as
\begin{align}\label{eq:GTN_ue_pool}
\nu_{m,k}=\frac{e^{\mathbf{c}^{\mathsf T}\mathbf{h}^{(L)}_{m,k}}}{\sum_{m'=1}^{M}e^{\mathbf{c}^{\mathsf T}\mathbf{h}^{(L)}_{m',k}}},
~~
\mathbf{g}^{\mathrm{UE}}_{k}=\sum_{m=1}^{M}\nu_{m,k}\mathbf{h}^{(L)}_{m,k}.
\end{align}

\subsection{Optimization Interface}\label{subsec:GTN_opt_interface}

To tightly couple the GTN with the optimization framework in Section \ref{sec:opt_solution}, we introduce a lightweight interface head that produces (1) a warm-start beamforming direction and (2) adaptive slack/penalty weights for feasibility control in SCA.

\subsubsection{GTN Warm-Start Beamformers}

We map each pair embedding $\mathbf{h}^{(L)}_{m,k}$ to an initial beam direction and normalize it to satisfy per-AP power budgets. To match the per-AP beamformer dimension $N_m$, we use an AP-specific projection $\mathbf{U}_{w,m}\in\mathbb{C}^{N_m\times d_h}$, which gives us
\begin{align}\label{eq:GTN_warm_start_0}
\widehat{\mathbf{w}}^{(0)}_{m,k}=\frac{\mathbf{U}_{w,m}\mathbf{h}^{(L)}_{m,k}}{\left\|\mathbf{U}_{w,m}\mathbf{h}^{(L)}_{m,k}\right\|+\epsilon_w},
\end{align}
leading to
\begin{align}\label{eq:GTN_warm_start}
\widehat{\mathbf{w}}^{(0)}_{m,k}\leftarrow
\widehat{\mathbf{w}}^{(0)}_{m,k}\sqrt{\frac{P_{\max}}{\sum_{k'=1}^{K}\left\|\widehat{\mathbf{w}}^{(0)}_{m,k'}\right\|^2+\epsilon_p}},
\end{align}
where $\leftarrow$ denotes an update operation, $\epsilon_w>0$ avoids division by zero, and $\epsilon_p>0$ stabilizes the per-AP renormalization, ensuring $\sum_{k}\|\widehat{\mathbf{w}}^{(0)}_{m,k}\|^2\le P_{\max}$ for each $m$. We then form $\widehat{\tilde{\mathbf{w}}}^{(0)}_{k}=\left[(\widehat{\mathbf{w}}^{(0)}_{1,k})^{\mathsf T},\ldots,(\widehat{\mathbf{w}}^{(0)}_{M,k})^{\mathsf T}\right]^{\mathsf T}$ and $\widehat{\mathbf{W}}^{(0)}_{k} = \widehat{\tilde{\mathbf{w}}}^{(0)}_{k}\big(\widehat{\tilde{\mathbf{w}}}^{(0)}_{k}\big)^{\mathsf H}$. To align with the initialization in Section \ref{sec:opt_solution}, we set $\delta_{m,k}^{(0)}=\xi_{m,k}$, unless otherwise specified. Using $\widehat{\mathbf{W}}^{(0)}=\{\widehat{\mathbf{W}}^{(0)}_{k}\}$ and $\boldsymbol{\delta}^{(0)}$, we compute $\zeta^{(0)}_{m,k}$ via \eqref{eq:GTN_sensing_cue} for B2S-consistent initialization.

\subsubsection{Adaptive Feasibility Weights}

To stabilize feasibility under tight SINR/CRB constraints, we introduce a nonnegative slack $\varsigma_k\ge0$ for each UE SINR constraint and weight it by a GTN-predicted difficulty score, given by
\begin{equation}\label{eq:GTN_lambda}
\lambda_k=\mathrm{softplus}\!\left(\mathbf{u}_{\lambda}^{\mathsf T}\mathbf{g}^{\mathrm{UE}}_{k}\right),
~~\varsigma_k\ge 0.
\end{equation}
Accordingly, the SINR constraint based on the $i$-th B2S iterate (cf. \eqref{eq:SINR_DC_compact_1}) is expressed as
\begin{equation}\label{eq:GTN_relaxed_sinr}
f_k(\mathbf{W};\boldsymbol{\delta}^{(i)})-g_k(\mathbf{W};\boldsymbol{\delta}^{(i)})+\varsigma_k
\geq\gamma_{\mathrm{th}}\left(I_k^{\mathrm{SI}}+\sigma^2\right).
\end{equation}
The slack is penalized in the optimization objective, e.g., by adding $\sum_{k}\lambda_k\varsigma_k$ to the beamforming subproblem objective \eqref{prob:W_subproblem}, where $\lambda_k$ assigns larger penalties to UEs whose cross-field geometry and visibility make feasibility harder. This couples GTN representations to the B2S trajectory while keeping the underlying solver unchanged. Thus, the GTN does not replace B2S, but improves its initialization, feasibility handling, and representation quality.

\section{Distributed Optimization-Learning with Graph Transformer Networks}\label{sec:DOLG}

The joint CF-ISAC beamforming problem in Section \ref{sec:II_system} is a mixed-integer non-convex program that tightly couples communication SINR, sensing CRB, per-AP power constraints, and cross near-/far-field propagation. While the B2S algorithm in Section \ref{sec:opt_solution} provides a principled optimization benchmark, its high computational complexity limits its scalability. To address this limitation, we propose DOLG, a distributed optimization-learning framework with GTN-conditioned MARL, that learns an optimization-aligned amortized surrogate of the B2S solution mapping and amortizes the optimization cost into offline MARL, enabling fast, non-iterative online execution.

\subsection{Optimization-Aligned DOLG Formulation}\label{subsec:DOLG_formulation}

In the proposed DOLG framework, each AP serves as a decentralized decision maker that selects local ISAC control variables at each epoch $t$ to jointly determine the system-wide communication and sensing performance. This can be viewed as an amortized approximation of \eqref{prob:P0}, where the iterative B2S policy is replaced by a one-shot, graph-conditioned decision rule that outputs relaxed AP-UE associations and beamformers consistent with the relaxation in Section \ref{sec:opt_solution}.

\subsubsection{State Representation}

The local observation available to the $m$-th AP is defined as
\begin{equation}\label{eq:DOLG_local_observation}
\mathbf{s}_m(t)
=
\big(
\mathbf{g}_m^{\mathrm{AP}}(t),
\boldsymbol{\psi}_m(t),
\boldsymbol{\iota}_m(t),
\boldsymbol{\varrho}_m(t)
\big),
\end{equation}
where $\mathbf{g}_m^{\mathrm{AP}}(t)$ denotes the AP-level graph embedding produced by the GTN described in Section \ref{sec:GTN_main}. This embedding captures cross-UE interference, cross-field propagation geometry, blockage visibility, and sensing relevance in a permutation-equivariant manner. The remaining components summarize local channel statistics, interference levels, and recent power usage. Once the GTN embedding is available at epoch $t$, each AP executes its policy using only its local observation tuple $\mathbf{s}_m(t)$. In \eqref{eq:DOLG_local_observation}, $\boldsymbol{\varrho}_m(t)$ avoids confusion with the prior uncertainty radius $\rho_s$ in Section \ref{sec:II_system}.

\subsubsection{Action Parameterization}

Each AP generates a continuous latent control variable $\mathbf{z}_m(t)$ from its local policy. The action is first mapped to (i) relaxed association coefficients and (ii) unconstrained beamformers via a structure-preserving transformation, given by
\begin{equation}\label{eq:DOLG_map}
\left(\{\tilde{\delta}_{m,k}(t)\}_k,\{\tilde{\mathbf{w}}_{m,k}(t)\}_k\right)=\mathcal{M}_{\mathrm{ISAC}}\left(\mathbf{z}_m(t)\right),
\end{equation}
which separates communication- and sensing-oriented components. To maintain consistency with the LoS-driven visibility mask in \eqref{eq:LOS_AP_set}--\eqref{prob:P0e}, we enforce the box constraint $0\le \delta_{m,k}(t)\le \xi_{m,k}$ via projection as
\begin{equation}
\delta_{m,k}(t)=\Pi_{[0,\xi_{m,k}]}\left(\tilde{\delta}_{m,k}(t)\right),~~\forall k,
\label{eq:DOLG_delta_proj}
\end{equation}
which yields a relaxed association rule aligned with the integer relaxation step in Section \ref{sec:opt_solution}.

To strictly enforce the per-AP transmit power constraint, we project $\{\tilde{\mathbf{w}}_{m,k}(t)\}_{k}$ onto the feasible set $\mathcal{P}_m$ defined by the per-AP power budget, which is given by
\begin{equation}\label{eq:DOLG_proj}
\begin{aligned}
\{\mathbf{w}_{m,k}(t)\}_{k=1}^{K}
&=\Pi_{\mathcal{P}_m}\left(\{\tilde{\mathbf{w}}_{m,k}(t)\}_{k=1}^{K}\right)\\
&\triangleq\arg\min_{\{\mathbf{w}_{m,k}\}}
\sum_{k=1}^{K}\left\|\mathbf{w}_{m,k}-\tilde{\mathbf{w}}_{m,k}(t)\right\|^2\\
\text{s.t.}&\sum_{k=1}^{K}\|\mathbf{w}_{m,k}\|^2\leq P_{\max},
\end{aligned}
\end{equation}
which guarantees feasibility with respect to the per-AP power constraint and enables non-iterative execution.

\begin{algorithm}[!t]
\caption{DOLG: Distributed Optimization-Learning with GTN}\label{alg:DOLG}
\small
\begin{algorithmic}[1]
\State \textbf{Input:}
Environment $\mathcal{E}_{\rm env}$; GTN encoder $\mathcal{F}_{\mathrm{GTN}}$;
actors $\{\pi_{\theta_m}\}$; critic $V_{\phi}$;
discount $\gamma_{\rm d}$; GAE $\lambda_{\rm GAE}$; PPO clip $\epsilon_{\rm clip}$;
weights $(\{\omega_k\},\omega_{\rm CRB},\omega_P)$.
\State \textbf{Initialize:}
$\{\theta_m\}$, $\phi$, rollout buffer $\mathcal{B}$.

\For{iteration $n=1,2,\ldots$}

\Statex \textbf{(A) Distributed rollout}
\For{$t=1$ to $T$}
    \State Sample network geometry and dynamics from $\mathcal{E}_{\rm env}$.
    \State Construct interaction graph $\mathcal{G}_t$ and encode
    $\{\mathbf{g}_m^{\mathrm{AP}}(t)\}_{m=1}^{M}=\mathcal{F}_{\mathrm{GTN}}(\mathcal{G}_t)$.
    \ForAll{{$m\in\mathcal{M}$} \textbf{in parallel}}
        \State Observe
        $\mathbf{s}_m(t)=(\mathbf{g}_m^{\mathrm{AP}}(t),\boldsymbol{\psi}_m(t),
        \boldsymbol{\iota}_m(t),\boldsymbol{\varrho}_m(t))$.
        \State Sample action
        $\mathbf{z}_m(t)\sim\pi_{\theta_m}(\cdot\mid\mathbf{s}_m(t))$.
        \State Map relaxed associations and beamformers.
        \State Project onto the feasible set
        $\delta_{m,k}(t)$.
    \EndFor
    \State Evaluate CF-ISAC metrics
    $\{R_k(t)\}$, $\{\gamma_k(t)\}$, $\{\mathrm{CRB}_s(t)\}$, and $P_{\mathrm{tot}}(t)$.
    \State Compute reward $r(t)$ and the stacked constraint residual vector
    $\boldsymbol{\Delta}(t)=
    \big[
    \{\Delta_k^{\mathrm{SINR}}(t)\}_{k}
    \ \|\ 
    \{\Delta_s^{\mathrm{CRB}}(t)\}_{s}
    \big]$.
    \State Store transition in $\mathcal{B}$.
\EndFor

\Statex \textbf{(B) Centralized training}
\State Compute returns and advantages using GAE.
\State Update critic $V_{\phi}$ by regressing on global inputs
$\big(\mathcal{G}_t,\boldsymbol{\Delta}(t)\big)$.
\ForAll{$m\in\mathcal{M}$}
    \State Update actor $\pi_{\theta_m}$ using PPO with clipping parameter $\epsilon_{\rm clip}$.
\EndFor
\State Clear buffer $\mathcal{B}$.

\EndFor
\end{algorithmic}
\end{algorithm}

\subsubsection{Reward Design}
To align DOLG with the optimization objective and constraints in \eqref{prob:P0}, we define a global scalar objective shared by all APs as
\begin{equation}
r(t)
=
\sum_{k=1}^{K}\omega_k R_k(t)
-
\omega_{\rm{CRB}}\sum_{s=1}^{S}\mathrm{CRB}_{s}(t)
-
\omega_{P} P_{\mathrm{tot}}(t),
\label{eq:DOLG_reward}
\end{equation}
where $R_k(t)$ and $\mathrm{CRB}_{s}(t)$ are evaluated according to the models in Section \ref{sec:II_system}. The objective in \eqref{eq:DOLG_reward} serves as an optimization-aligned surrogate of \eqref{prob:P0} and can be interpreted as a Lagrangian-type surrogate with fixed multipliers $(\omega_{\rm{CRB}},\omega_{P})$.

\subsubsection{Transition Dynamics}
State transitions are governed by the underlying THz CF-ISAC environment, including cross-field propagation, near-/far-field regime switching, UE mobility, blockage dynamics, and ISAC waveform interaction. These dynamics determine the evolution of the interaction graph $\mathcal{G}_t$ and the resulting GTN embeddings across time.

\subsection{Constraint-Aware Environment Feedback}\label{subsec:DOLG_feedback}

Beyond the scalar objective \eqref{eq:DOLG_reward}, the environment provides structured feedback that mirrors the constraint set of the original formulation. After executing $\{\delta_{m,k}(t),\mathbf{w}_{m,k}(t)\}$ (and fixed sensing pilots), we compute the SINR constraint margin as
\begin{equation}\label{eq:DOLG_sinr_margin}
\Delta_k^{\mathrm{SINR}}(t)
\triangleq
\gamma_k(t)-\gamma_{\mathrm{th}},
~~ \forall k\in\mathcal{K},
\end{equation}
and the CRB constraint margin as
\begin{equation}\label{eq:DOLG_crb_margin}
\Delta_{s}^{\mathrm{CRB}}(t)
\triangleq
\epsilon_{\mathrm{th}}-\mathrm{CRB}_{s}(t),
~~ \forall s\in\mathcal{S}.
\end{equation}
Then, we stack the residuals as
\begin{equation}\label{eq:DOLG_delta}
\boldsymbol{\Delta}(t)\triangleq\left[\{\Delta_k^{\mathrm{SINR}}(t)\}_{k} \ \|\ \{\Delta_{s}^{\mathrm{CRB}}(t)\}_{s}\right],
\end{equation}
which serves as the optimization-aligned side information during training.

\subsection{GTN-Conditioned DOLG Algorithm}\label{subsec:DOLG_algorithm}

DOLG is implemented as a GTN-conditioned MARL framework under CTDE. At each decision epoch, the GTN encodes the heterogeneous interaction graph to generate AP embeddings, based on which each AP outputs a latent control variable that is mapped to relaxed associations and ISAC beamformers via structure-preserving transformations and the feasibility projection in \eqref{eq:DOLG_proj}. The environment evaluates the resulting joint action using the exact THz CF-ISAC model. During training, a centralized critic and PPO are used to stabilize learning under inter-AP coupling, while during execution each AP acts independently based on its local observation and GTN embedding.
DOLG can be interpreted as an amortized, learning based approximation to the optimization based B2S solver for CF-ISAC beamforming. Let $\left(\boldsymbol{\delta}^{\star}(t),\mathbf{W}^{\star}(t)\right)$ denote the B2S solution for the realized environment at epoch $t$. Instead of iteratively solving convexified subproblems, DOLG produces a one-shot approximation, given by
\begin{align}\label{eq:DOLG_amortized}
\left(\widehat{\boldsymbol{\delta}}(t),\widehat{\mathbf{W}}(t)\right)&=
\mathcal{A}_{\theta}\left(\mathcal{G}_t,\{\boldsymbol{\psi}_m(t),\boldsymbol{\iota}_m(t),\boldsymbol{\varrho}_m(t)\}_m\right)\notag\\
&\approx\left(\boldsymbol{\delta}^{\star}(t),\mathbf{W}^{\star}(t)\right),
\end{align}
where $\mathcal{A}_{\theta}$ denotes the GTN-conditioned policy pipeline.
From the optimization perspective, the MARL formulation underlying DOLG is the original non-convex ISAC problem, where the reward in \eqref{eq:DOLG_reward} acts as a Lagrangian surrogate and the feasibility projection in \eqref{eq:DOLG_proj} enforces hard per-AP power constraints. The resulting objective and constraint residual feedback guide the learned policy towards optimization-consistent solutions. Meanwhile, the GTN embeddings capture cross-field geometry, multi-AP coupling, and sensing relevance, enabling centralized training to approximate centralized optimization while decentralized execution preserves scalability. Together, DOLG forms a coherent \emph{optimization-representation-learning} pipeline for large-scale CF-ISAC beamforming. For clarity, the overall GTN-conditioned DOLG procedure is summarized in Algorithm~\ref{alg:DOLG}.

\subsection{Computational Complexity Analysis}\label{subsec:complexity}

We now compare the computational complexity of the optimization based B2S benchmark, heuristic beamforming schemes, and the proposed DOLG framework. For clarity, we assume $N_m = N$ antennas at each AP and $N_{\mathrm{tot}} = MN$.

\subsubsection{B2S Benchmark}
In Section \ref{sec:opt_solution}, the SDR lifting yields $K$ Hermitian covariances $\{\mathbf{W}_k\succeq\mathbf{0}\}_{k\in\mathcal{K}}$ with $\mathbf{W}_k\in\mathbb{C}^{N_{\mathrm{tot}}\times N_{\mathrm{tot}}}$. Each BCD iteration solves an SDP over these $K$ matrices under semidefinite and affine constraints. Using interior-point methods, the worst-case per-iteration complexity scales as $\mathcal{C}_{\mathrm{B2S}}=\tilde{\mathcal{O}}\left(I_{\mathrm{BCD}}\mathrm{poly}(K)N_{\mathrm{tot}}^{\alpha}
\right),~~\alpha>0,$
where $\mathrm{poly}(K)$ denotes a polynomial function of $K$, which grows rapidly with $M$ through $N_{\mathrm{tot}}=MN$ and motivates amortized learning for large-scale deployments.

\subsubsection{Heuristic Baselines}
We note that MRT and ZF-type beamforming schemes incur complexities of $\mathcal{O}(MKN)$ and $\mathcal{O}(MK^3 + MKN^2)$, respectively. Despite their computational efficiency, these schemes fail to enforce sensing constraints or capture cross-field ISAC coupling.

\subsubsection{Proposed DOLG}
The computational burden is shifted to offline training. At runtime, DOLG performs a single GTN forward pass and one actor evaluation, leading to $\mathcal{C}_{\mathrm{online}}=\mathcal{O}\left(L_g (H|\mathcal{E}| d + MK d^2) + M d^2\right),$ which is non-iterative and approximately linear in $M$ under practical sparse-visibility THz graphs. Overall, DOLG achieves a favorable complexity-performance tradeoff by amortizing the optimization cost into offline learning while enabling fast, distributed online execution.

\section{Simulation Results and Discussion}\label{sec:simulation_results}

In this section, we evaluate the performance of the proposed DOLG framework under practical cross-field propagation while examining the impacts of blockage and multi-AP interference, based on the THz communication and sensing models in Section \ref{sec:II_system}. Specifically, we adopt a dual-perspective evaluation: (i) a system-design perspective comparing different transmission architectures, and (ii) an algorithmic perspective comparing DOLG against B2S, heuristic, and MARL without GTN under the same CF-ISAC design. The performance is measured in terms of sensing accuracy (i.e., CRB), energy efficiency, computational complexity, and learning convergence.


Our evaluation focuses on the downlink of a THz CF-ISAC system with $M$ distributed APs, $K$ single-antenna UEs, and $S$ STs randomly deployed in a $100\times100~\mathrm{m}^2$ area. Unless otherwise specified, APs, UEs, and STs are independently and uniformly distributed, and all results are averaged over $200$ random realizations. Each AP is equipped with a ULA with $N_m=32$ antennas and half-wavelength spacing. The carrier frequency and bandwidth are set to $f_c=0.3~\mathrm{THz}$ and $B=5~\mathrm{GHz}$, respectively. The THz channel follows the model in Section~\ref{sec:II_system}, where path loss includes both free-space attenuation and molecular absorption as in \eqref{eq:THz_pathloss}, and the array response follows the unified cross-field model as in \eqref{eq:unified_response} with coexisting near- and far-field links determined by the Rayleigh distance. Blockage is modeled by the distance-dependent LoS probability $p_{\mathrm{LoS}}(r)=e^{-\beta r}$ with $\beta=0.01$, and the candidate AP set for each UE is determined as in \eqref{eq:LOS_AP_set} with $p_{\mathrm{th}}=0.5$, yielding a sparse and visibility-aware topology. The maximum transmit power per AP is $P_{\max}=30~\mathrm{dBm}$, the noise power is calculated as $\sigma^2=-174~\mathrm{dBm/Hz}+10\log_{10}(B)+F$ with noise figure $F=7~\mathrm{dB}$ \cite{Bjornson2017Massive}, the SINR threshold is $\gamma_{\mathrm{th}}=5~\mathrm{dB}$, and the sensing accuracy requirement is $\epsilon_{\mathrm{th}}=10^{-2}$. 
For sensing, the system estimates the ST parameters through the echo model in Section~\ref{sec:II_system}, and the sensing accuracy is quantified by the CRB in \eqref{eq:crb}; for communication, the achievable rate is evaluated based on the received SINR in \eqref{eq:SINR_k}. 
The B2S algorithm is implemented in CVX with convergence tolerance $10^{-3}$ and a maximum of $50$ iterations. The proposed DOLG framework adopts centralized training and decentralized execution, where each AP is treated as an agent and the policy is trained with a learning rate of $3\times10^{-4}$, a discount factor of $0.99$, and a batch size of $1024$. The GTN uses $L=3$ layers, each featuring $4$ attention heads, with a hidden dimension of $d_h=128$. All compared schemes are evaluated under identical random network realizations and system parameters to ensure fair comparison and reproducibility.

\subsection{Performance from System-design Perspective}\label{subsec:system_design_results}

\begin{figure}[!t]
\centering
\includegraphics[height=2.4in,width=0.95\columnwidth]{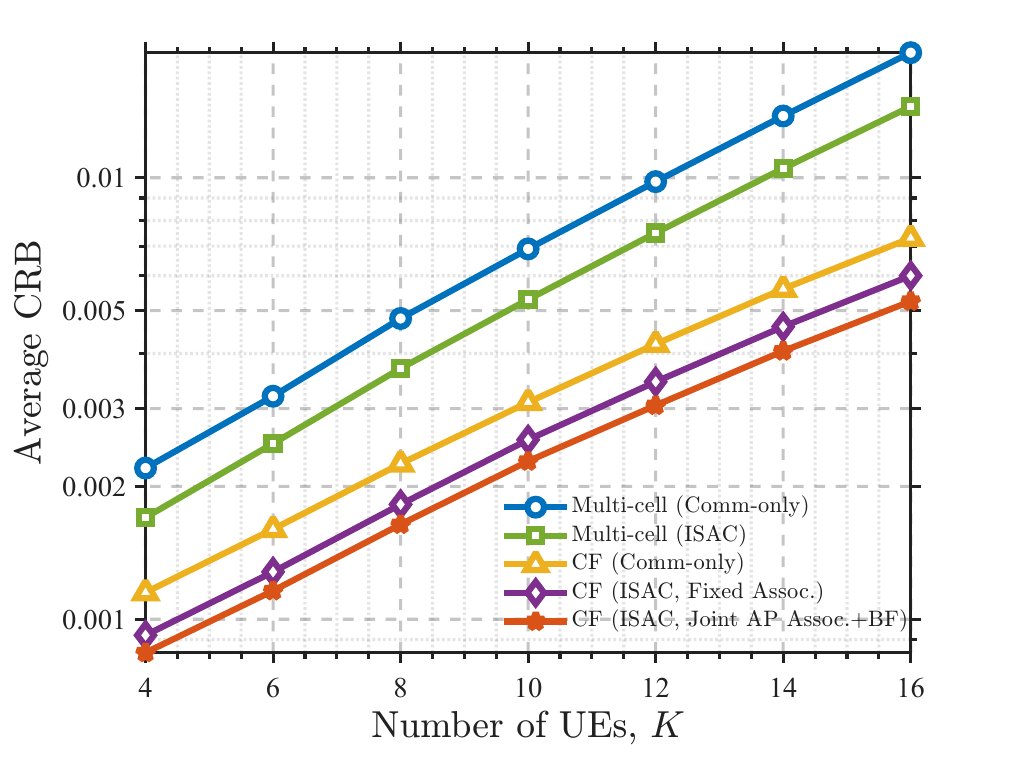}
\vspace{-8pt}
\caption{Average CRB versus the number of UEs for different schemes.}
\vspace{-8pt}
\label{fig:crb_comparison}
\end{figure}

\begin{figure}[!t]
\centering
\includegraphics[height=2.4in,width=0.95\columnwidth]{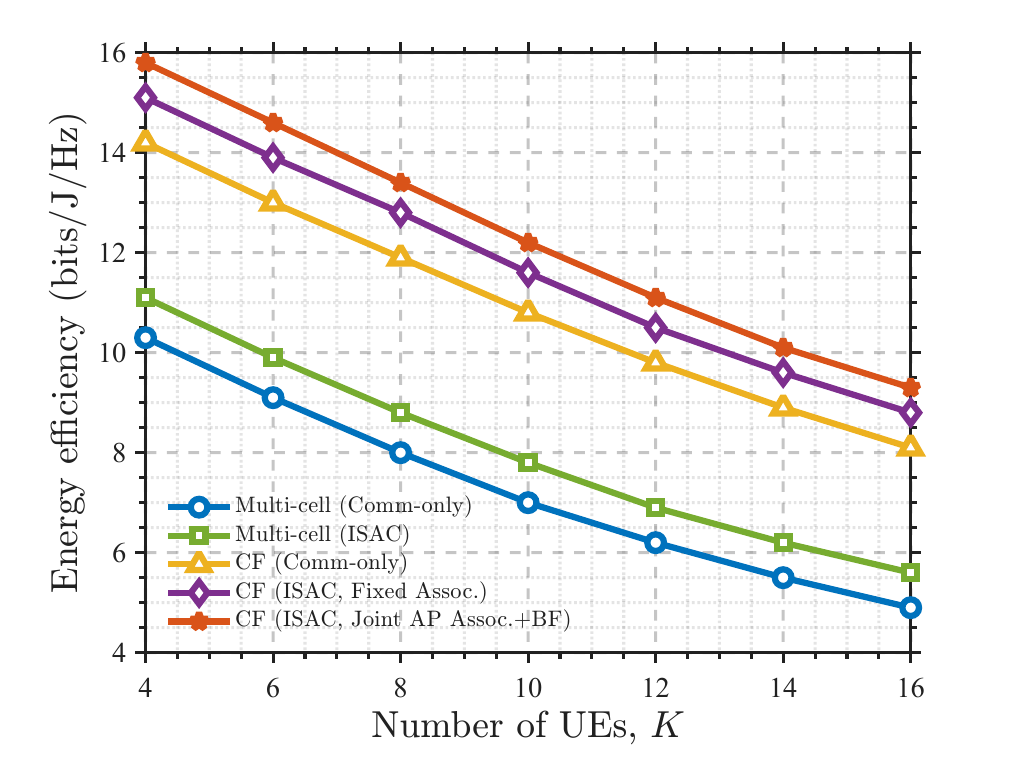}
\vspace{-8pt}
\caption{Energy efficiency versus the number of UEs for different schemes.}
\vspace{-8pt}
\label{fig:ee_comparison}
\end{figure}

We first compare different transmission architectures from the system-design perspective. Specifically, the compared schemes include: 1) multi-cell communication-only design, 2) multi-cell ISAC design, 3) CF communication-only design, 4) CF-ISAC design with fixed association, and 5) CF-ISAC design with joint AP association and beamforming. For the communication-only baselines, the sensing pilots are fixed \textit{a priori}, while only the communication transmission variables are optimized. The resulting transmit signals are then substituted into the same sensing echo and CRB models in Section \ref{sec:II_system}, such that all schemes are evaluated under a unified sensing-performance metric. To ensure a fair system-level comparison, all five schemes are implemented and evaluated using the optimization based B2S benchmark.


Fig. \ref{fig:crb_comparison} illustrates the average CRB achieved by five schemes versus the number of UEs, $K$. As $K$ increases, the CRB achieved by all schemes increases monotonically. Notably, the CF-ISAC design with joint AP association and beamforming consistently achieves the lowest CRB, followed by the CF-ISAC design with fixed association and the CF communication-only design. In the multi-cell architecture, the ISAC design also yields a lower CRB compared with its communication-only counterpart; however, both remain noticeably inferior to the CF schemes. These results indicate that the sensing performance improves in two progressive steps: first, transitioning from communication-only design to ISAC design reduces the CRB in both multi-cell and CF architectures; second, migrating from multi-cell transmission to CF cooperative transmission further reduces the CRB. Furthermore, the joint optimization of AP association and beamforming provides an extra gain over the ISAC design with fixed association.


Fig. \ref{fig:ee_comparison} compares the energy efficiency of five schemes versus the number of UEs, $K$. For all schemes, energy efficiency decreases monotonically as $K$ increases, since a larger user load requires more transmit power and leads to a more interference-limited communication environment. The CF-ISAC design with joint AP association and beamforming maintains a clear lead, achieving the highest energy efficiency, followed by the CF-ISAC design with fixed association and the CF communication-only design. In the multi-cell architecture, the ISAC design also outperforms the communication-only counterpart, although both are markedly below the CF schemes. This underscores an efficiency gain achieved by moving from communication-only design to ISAC design and moving from multi-cell transmission to CF cooperative transmission. Ultimately, the superior performance of the joint optimization of AP association and beamforming demonstrates its benefits on the most effective utilization of communication and sensing resources.

\vspace{-8pt}
\subsection{Performance from Algorithmic Perspective}\label{subsec:algorithmic_results}

\begin{figure}[t]
    \centering
    \subfloat[\label{fig:algorithmic_comparison_a}]{\includegraphics[height=2.4in,width=0.95\columnwidth]{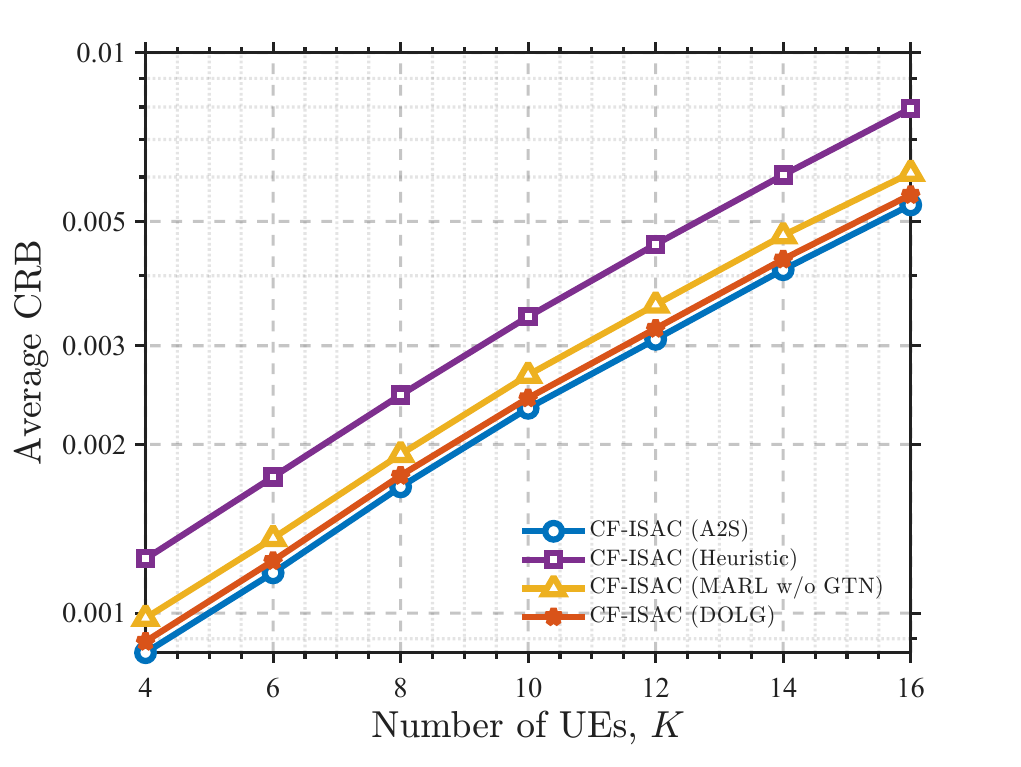}}\\[-2pt]
    \subfloat[\label{fig:algorithmic_comparison_b}]{\includegraphics[height=2.4in,width=0.95\columnwidth]{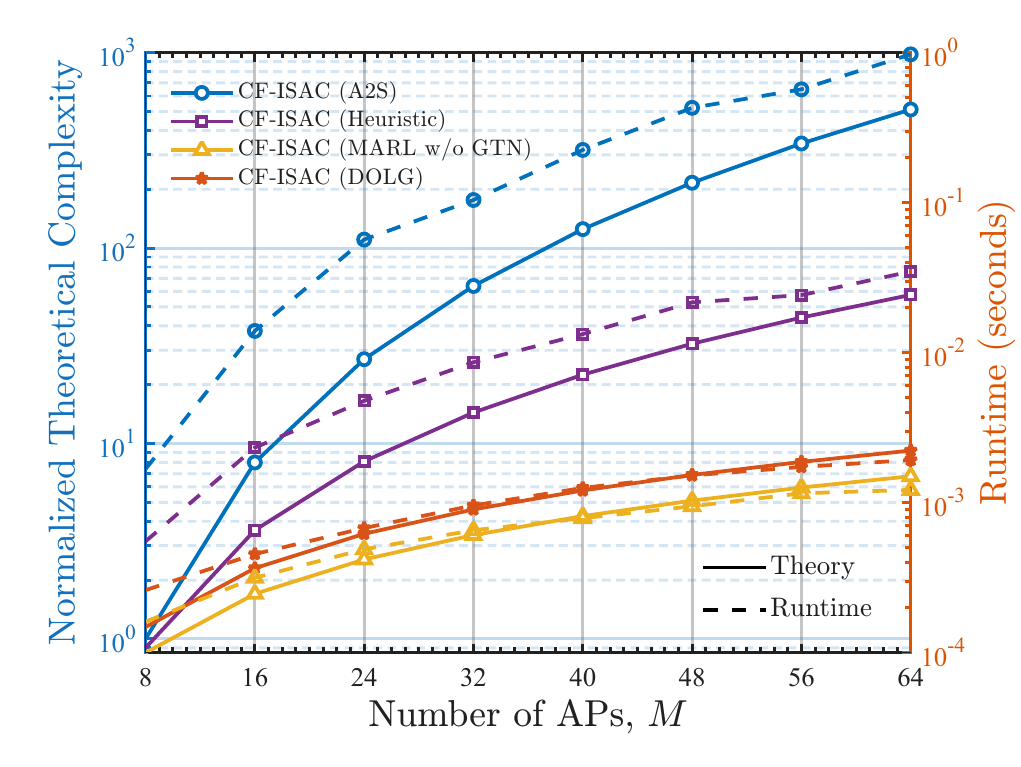}}
    \caption{Algorithmic comparison of different solutions for the CF-ISAC design.}
    \vspace{-8pt}
    \label{fig:algorithmic_comparison}
\end{figure}

Next, we evaluate four algorithmic solutions, including the optimization based B2S benchmark, a low-complexity heuristic design, a MARL baseline without the GTN encoder, and the proposed DOLG framework, using the same CF-ISAC design with the joint AP association and beamforming introduced in the system-design comparison. As shown in Fig. \ref{fig:algorithmic_comparison}(a), the average CRB for all solutions increases with the number of UEs, $K$, while the relative performance ranking remains consistent, i.e., B2S achieves the lowest CRB, closely followed by DOLG, while MARL without GTN and the heuristic design exhibit noticeably higher sensing errors. Fig. \ref{fig:algorithmic_comparison}(b) illustrates the theoretical complexity (left axis) and empirical runtime (right axis) versus the number of APs, $M$. As predicted in Section \ref{subsec:complexity}, the runtime of B2S escalates rapidly with $M$ due to the repeated solution of semidefinite subproblems. Conversely, the heuristic, MARL without GTN, and DOLG solutions maintain significantly lower runtimes. Although the GTN encoder introduces a slight computational overhead compared to plain MARL, DOLG remains far more efficient than B2S while achieving more accurate sensing performance. Consequently, DOLG offers a superior performance-complexity tradeoff for large-scale THz CF-ISAC systems.


\vspace{-8pt}
\subsection{Training Behavior of Proposed DOLG Framework}\label{subsec:training_behavior}

\begin{figure*}[t]
    \centering
    \subfloat[Objective convergence\label{fig:dolg_obj}]{
        \includegraphics[width=0.3\textwidth,height=1.6in]{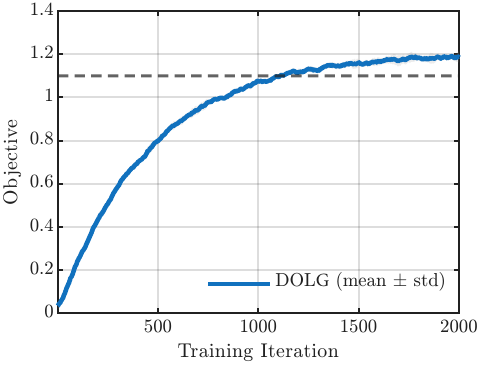}
    }\hfill
    \subfloat[CRB and sum-rate evolution\label{fig:dolg_crb_rate}]{
        \includegraphics[width=0.32\textwidth,height=1.6in]{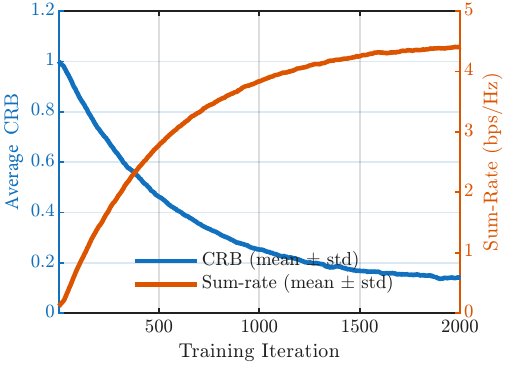}
    }\hfill
    \subfloat[Constraint violation\label{fig:dolg_viol}]{
        \includegraphics[width=0.3\textwidth,height=1.6in]{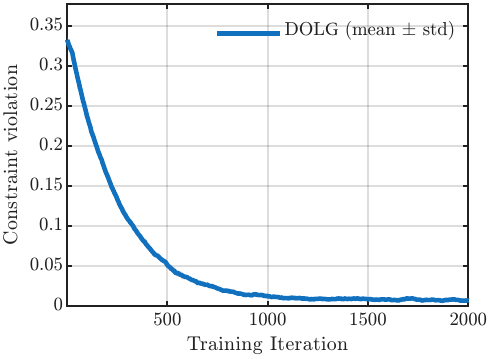}
    }
    \caption{Training behavior of the proposed DOLG framework.}
    \vspace{-12pt}
    \label{fig:dolg_training}
\end{figure*}

Finally, we investigate the training behavior of the DOLG framework. Fig. \ref{fig:dolg_training} illustrates the mean learning curves across multiple independent runs, with shaded regions representing one standard deviation. As shown in Fig. \ref{fig:dolg_training}(a), the optimization-aligned objective exhibits rapid gains during the initial training phase before reaching a steady state, validating that the proposed GTN-conditioned MARL framework effectively learns a distributed policy for THz CF-ISAC systems. Fig. \ref{fig:dolg_training}(b) depicts the simultaneous increase in communication sum-rate and reduction in sensing CRB, confirming that the learned policy progressively improves the communication-sensing tradeoff. Furthermore, Fig. \ref{fig:dolg_training}(c) shows that constraint violations diminish rapidly and remain negligible after convergence. Overall, these results demonstrate that DOLG not only achieves superior online performance but also exhibits robust and stable learning dynamics.

\vspace{-10pt}
\section{Conclusion} \label{sec:VII_conclusion}

In this paper, we proposed a distributed optimization--learning framework for THz CF-ISAC systems. By addressing a highly non-convex joint scheduling and signal design problem under cross-field propagation and sensing constraints, we developed a GTN-conditioned multi-agent architecture that converts optimization-guided design into a scalable one-shot distributed policy. Simulation results showed that the proposed framework achieves stable convergence while effectively balancing communication and sensing performance. Furthermore, the framework maintains near-linear online complexity relative to system size and consistently outperforms conventional optimization based, heuristic, and non-joint design baselines. These findings demonstrate the potential of optimization-aligned learning for scalable and adaptive THz CF-ISAC deployments.

\appendices

\section{Derivation of CRB Matrix}\label{App:Derivation_CRB_Matrix}

This appendix derives the FIM and the corresponding CRB under the stacked multi-AP sensing echo model. 
Let $\mathbf{G}_{s}(\boldsymbol{\xi}_s)\triangleq
\mathrm{blkdiag}\big(\mathbf{G}_{1,s}(\boldsymbol{\xi}_s),\ldots,\mathbf{G}_{M,s}(\boldsymbol{\xi}_s)\big)$ denote the stacked round-trip channel matrix. At time slot $t$, we define the stacked transmit vector as $\mathbf{x}[t]\triangleq[\mathbf{x}_{1}^{\mathsf T}[t],\ldots,\mathbf{x}_{M}^{\mathsf T}[t]]^{\mathsf T}$ and collect the transmit signals over $T$ slots as
$\mathbf{X}\triangleq[\mathbf{x}[1],\ldots,\mathbf{x}[T]]$. Stacking the multi-AP echo signals over $T$ slots yields $\mathbf{Y}^{\mathrm{echo}}_s=\mathbf{G}_{s}(\boldsymbol{\xi}_s)\mathbf{X}+\mathbf{Z}$, where $\mathbf{Z}$ collects the sensing noise samples and satisfies $\mathrm{vec}(\mathbf{Z})\sim\mathcal{CN}(\mathbf{0},N_0\mathbf{I})$. We denote $\mathbf{Y}^{\mathrm{vec}}_s \triangleq \mathrm{vec}(\mathbf{Y}^{\mathrm{echo}}_s)$ as the vectorized stacked echo observations and define $\mathbf{x}_{\mathrm{vec}}\triangleq\mathrm{vec}(\mathbf{X})$. Since $\mathrm{vec}(\mathbf{Z})\sim\mathcal{CN}(\mathbf{0},N_0\mathbf{I})$, we find that $\mathbf{Y}^{\mathrm{vec}}_s\sim\mathcal{CN}\!\big(\boldsymbol{\Upsilon}(\boldsymbol{\xi}_s),N_0\mathbf{I}\big)$, where the mean vector is given by
\begin{equation}\label{eq:mean_vec}
\boldsymbol{\Upsilon}(\boldsymbol{\xi}_s)=(\mathbf{I}_T\otimes\mathbf{G}_{s}(\boldsymbol{\xi}_s))\,\mathbf{x}_{\mathrm{vec}},
\end{equation}
where $\otimes$ denotes the Kronecker product. Hence, for complex Gaussian observations whose mean depends on $\boldsymbol{\xi}_s$ only, the FIM is obtained as
\begin{equation}\label{eq:FIM_block}
\mathbf{J}_{\boldsymbol{\xi}_s}=\frac{2}{N_0}\operatorname{Re}
\left\{\Big(\frac{\partial\boldsymbol{\Upsilon}}{\partial\boldsymbol{\xi}_s} \Big)^{{\mathsf H}}
\Big(\frac{\partial\boldsymbol{\Upsilon}}{\partial\boldsymbol{\xi}_s}\Big)\right\}=
\begin{bmatrix}
\mathbf{J}_{11}&\mathbf{J}_{12}\\
\mathbf{J}_{12}^{\mathsf H}&\mathbf{J}_{22}
\end{bmatrix}.
\end{equation}

\subsection*{I. Jacobians w.r.t.\ Parameters}

We use the same composite parameter vector as in the main text, i.e., $\boldsymbol{\xi}_s=[\boldsymbol{\xi}_{1,s}^{\mathsf T},\ldots,\boldsymbol{\xi}_{M,s}^{\mathsf T}]^{\mathsf T}$,
where $\boldsymbol{\xi}_{m,s}=[\,r_{m,s},\ \theta_{m,s},\ \operatorname{Re}(\beta_{m,s}^{\mathrm{rt}}),\ \operatorname{Im}(\beta_{m,s}^{\mathrm{rt}})\,]^{\mathsf T}$
and $\beta_{m,s}^{\mathrm{rt}}=\operatorname{Re}(\beta_{m,s}^{\mathrm{rt}})+j\operatorname{Im}(\beta_{m,s}^{\mathrm{rt}})$.
For brevity, we define $\mathbf{a}\triangleq\mathbf{a}(\theta_{m,s},r_{m,s})$, as well as the array-derivative vectors as 
${\mathbf{a}}_{r_{m,s}}\triangleq\partial\mathbf{a}(\theta_{m,s},r_{m,s})/\partial r_{m,s}$ and
${\mathbf{a}}_{\theta_{m,s}}\triangleq\partial\mathbf{a}(\theta_{m,s},r_{m,s})/\partial\theta_{m,s}$.
As per the product rule, we obtain
\begin{subequations}\label{eq:dG_all}
\begin{align}
\frac{\partial\mathbf{G}_{m,s}}{\partial r_{m,s}}
&=\beta_{m,s}^{\mathrm{rt}}\big(\mathbf{a}_{r_{m,s}}\,\mathbf{a}^{\mathsf T} + \mathbf{a}\,\mathbf{a}_{r_{m,s}}^{\mathsf T}\big)
\triangleq \mathbf{G}_{r_{m,s}},
\label{eq:dG_dr}\\
\frac{\partial\mathbf{G}_{m,s}}{\partial \theta_{m,s}}
&=\beta_{m,s}^{\mathrm{rt}}\big(\mathbf{a}_{\theta_{m,s}}\,\mathbf{a}^{\mathsf T} + \mathbf{a}\,\mathbf{a}_{\theta_{m,s}}^{\mathsf T}\big)
\triangleq \mathbf{G}_{\theta_{m,s}},
\label{eq:dG_dtheta}\\
\frac{\partial\mathbf{G}_{m,s}}{\partial \operatorname{Re}(\beta_{m,s}^{\mathrm{rt}})}
&=\mathbf{a}\mathbf{a}^{\mathsf T},
\label{eq:dG_dRebeta}\\
\frac{\partial\mathbf{G}_{m,s}}{\partial \operatorname{Im}(\beta_{m,s}^{\mathrm{rt}})}
&=j\,\mathbf{a}\mathbf{a}^{\mathsf T}.
\label{eq:dG_dImbeta}
\end{align}
\end{subequations}
To keep the stacked notation compact, we define the embedded derivative block as
\begin{equation}\label{eq:embedded_G_def}
\widetilde{\mathbf{G}}_{p_m}
\triangleq
\mathrm{blkdiag}(\mathbf{0},\ldots,\mathbf{0},\mathbf{G}_{p_m},\mathbf{0},\ldots,\mathbf{0}),
\end{equation}
where $p_m\in\{r_{m,s},\theta_{m,s},\operatorname{Re}(\beta_{m,s}^{\mathrm{rt}}),\operatorname{Im}(\beta_{m,s}^{\mathrm{rt}})\}$, and the only nonzero block is placed at the $m$-th AP position.  In particular, we find that $\mathbf{G}_{p_m}=\mathbf{G}_{r_{m,s}}$ and $\mathbf{G}_{p_m}=\mathbf{G}_{\theta_{m,s}}$ for $p_m=r_{m,s}$ and $p_m=\theta_{m,s}$, respectively. For $p_m\in\{\operatorname{Re}(\beta_{m,s}^{\mathrm{rt}}),\operatorname{Im}(\beta_{m,s}^{\mathrm{rt}})\}$, the corresponding blocks are $\mathbf{a}\mathbf{a}^{\mathsf T}$ and $j\mathbf{a}\mathbf{a}^{\mathsf T}$, respectively. Using \eqref{eq:mean_vec}, the Jacobian column corresponding to $p_m$ is obtained as
\begin{align}\label{eq:partials_mu}
\frac{\partial\boldsymbol{\Upsilon}}{\partial p_m}=(\mathbf{I}_T\otimes \widetilde{\mathbf{G}}_{p_m})\,\mathbf{x}_{\mathrm{vec}}.
\end{align}

\subsection*{II. From Quadratic Forms to Trace Forms}

For the $m$-th AP, we define the per-AP transmit covariance as $\mathbb{X}_m\triangleq \frac{1}{T}\sum_{t=1}^{T}\mathbf{x}_m[t]\mathbf{x}_m^{\mathsf H}[t]$. For any $p_m,q_m\in\{r_{m,s},\theta_{m,s}\}$ associated with the $m$-th AP, we obtain
\begin{align}\label{eq:Jlp_trace0}
J_{p_m q_m}&=\frac{2}{N_0}\operatorname{Re}\left\{\mathbf{x}_{\mathrm{vec}}^{\mathsf H}(\mathbf{I}_T\otimes \widetilde{\mathbf{G}}_{p_m}^{\mathsf H})
(\mathbf{I}_T\otimes \widetilde{\mathbf{G}}_{q_m})\mathbf{x}_{\mathrm{vec}}\right\} \nonumber\\
&=\frac{2}{N_0}\operatorname{Re}\left\{\sum_{t=1}^{T}\mathbf{x}_m[t]^{\mathsf H}{\mathbf{G}}_{p_m}^{\mathsf H}{\mathbf{G}}_{q_m}\,\mathbf{x}_m[t]\right\}\nonumber\\
&=\frac{2T}{N_0}\operatorname{Re}\left\{\operatorname{tr}\left({\mathbf{G}}_{q_m}\,\mathbb{X}_m\,{\mathbf{G}}_{p_m}^{\mathsf H}\right)\right\}.
\end{align}
For the parameters associated with other APs, e.g., the $n$-th AP with $n\neq m$, the corresponding FIM entries are zero due to the block-diagonal structure of $\mathbf{G}_s(\boldsymbol{\xi}_s)$, i.e., $J_{p_m q_n}=0$. The cross terms with the reflection-coefficient parameters follow from \eqref{eq:dG_all} and \eqref{eq:partials_mu}. For $p_m\in\{r_{m,s},\theta_{m,s}\}$, we obtain
\begin{subequations}\label{eq:J_l_beta}
\begin{align}
J_{p_m,\operatorname{Re}(\beta_{m,s}^{\mathrm{rt}})}
&=\frac{2T}{N_0}\operatorname{Re}\!\left\{\operatorname{tr}
\big((\mathbf{a}\mathbf{a}^{\mathsf T})\,\mathbb{X}_m\,\mathbf{G}_{p_m}^{\mathsf H}\big)\right\},
\label{eq:J_l_beta_a}\\
J_{p_m,\operatorname{Im}(\beta_{m,s}^{\mathrm{rt}})}
&=\frac{2T}{N_0}\operatorname{Re}\!\left\{\operatorname{tr}
\big((j\mathbf{a}\mathbf{a}^{\mathsf T})\,\mathbb{X}_m\,\mathbf{G}_{p_m}^{\mathsf H}\big)\right\}.
\label{eq:J_l_beta_b}
\end{align}
\end{subequations}
For the $(\operatorname{Re}(\beta_{m,s}^{\mathrm{rt}}),\operatorname{Im}(\beta_{m,s}^{\mathrm{rt}}))$ block, we obtain
\begin{subequations}\label{eq:J_beta_beta}
\begin{align}
J_{\operatorname{Re}(\beta_{m,s}^{\mathrm{rt}}),\operatorname{Re}(\beta_{m,s}^{\mathrm{rt}})}
&=\frac{2T}{N_0}\operatorname{tr}\!\big((\mathbf{a}\mathbf{a}^{\mathsf T})\,\mathbb{X}_m\,(\mathbf{a}\mathbf{a}^{\mathsf T})^{\mathsf H}\big),
\label{eq:J_beta_beta_a}\\
J_{\operatorname{Im}(\beta_{m,s}^{\mathrm{rt}}),\operatorname{Im}(\beta_{m,s}^{\mathrm{rt}})}
&=J_{\operatorname{Re}(\beta_{m,s}^{\mathrm{rt}}),\operatorname{Re}(\beta_{m,s}^{\mathrm{rt}})},
\label{eq:J_beta_beta_b}\\
J_{\operatorname{Re}(\beta_{m,s}^{\mathrm{rt}}),\operatorname{Im}(\beta_{m,s}^{\mathrm{rt}})}
&=0.
\label{eq:J_beta_beta_c}
\end{align}
\end{subequations}

\subsection*{III. Block Structure and CRB}

Collecting the afore-derived entries, the FIM for the $s$-th ST admits a block-diagonal structure across APs, given by
\begin{equation}\label{eq:FIM_blkdiag_APs_app}
\mathbf{J}_{\boldsymbol{\xi}_s}=\mathrm{blkdiag}\{\mathbf{J}_{\boldsymbol{\xi}_{1,s}},\ldots,\mathbf{J}_{\boldsymbol{\xi}_{M,s}}\}.
\end{equation}
Accordingly, each per-AP FIM block $\mathbf{J}_{\boldsymbol{\xi}_{m,s}}\in\mathbb{C}^{4\times 4}$ is partitioned as
\begin{equation}\label{eq:FIM_block_m_app}
\mathbf{J}_{\boldsymbol{\xi}_{m,s}}=\begin{bmatrix}
\mathbf{J}_{11,m,s}&\mathbf{J}_{12,m,s}\\
\mathbf{J}_{12,m,s}^{\mathsf H}&\mathbf{J}_{22,m,s}
\end{bmatrix},
\end{equation}
where $\mathbf{J}_{11,m,s}\in\mathbb{R}^{2\times 2}$ is the FIM submatrix with respect to the geometric parameters $(r_{m,s},\theta_{m,s})$,
$\mathbf{J}_{22,m,s}\in\mathbb{R}^{2\times 2}$ is the FIM submatrix with respect to the reflection-coefficient parameters $(\operatorname{Re}(\beta_{m,s}^{\mathrm{rt}}),\operatorname{Im}(\beta_{m,s}^{\mathrm{rt}}))$,
and $\mathbf{J}_{12,m,s}\in\mathbb{R}^{2\times 2}$ collects the cross-information between these two parameter groups.
Here, the entries of $\mathbf{J}_{11,m,s}$, $\mathbf{J}_{12,m,s}$, and $\mathbf{J}_{22,m,s}$ follow the same definitions as in \eqref{eq:FIM_J11}--\eqref{eq:FIM_J22}. Furthermore, we define the stacked blocks for the $s$-th ST as
\begin{equation}\label{eq:stacked_J_blocks_s_app}
\mathbf{J}_{ij,s}\triangleq\mathrm{blkdiag}\{\mathbf{J}_{ij,1,s},\ldots,\mathbf{J}_{ij,M,s}\}
\end{equation}
with $(i,j)\in\{(1,1),(1,2),(2,2)\}$. Then, \eqref{eq:FIM_blkdiag_APs_app} can be equivalently written as
\begin{equation}\label{eq:FIM_block_form_s_app}
\mathbf{J}_{\boldsymbol{\xi}_s}
=\begin{bmatrix}
\mathbf{J}_{11,s}&\mathbf{J}_{12,s}\\
\mathbf{J}_{12,s}^{\mathsf H}&\mathbf{J}_{22,s}
\end{bmatrix}.
\end{equation}

If the sensing echoes of different STs are statistically independent under the adopted stacked echo model, then the global FIM for all STs is block-diagonal across $s$, i.e.,
\begin{equation}\label{eq:FIM_global_blkdiag_STs_app}
\mathbf{J}_{\boldsymbol{\xi}}=\mathrm{blkdiag}\{\mathbf{J}_{\boldsymbol{\xi}_{1}},\ldots,\mathbf{J}_{\boldsymbol{\xi}_{S}}\}.
\end{equation}
Its corresponding sub-blocks are given by
\begin{equation}\label{eq:global_subblocks_app}
\mathbf{J}_{ij}\triangleq\mathrm{blkdiag}\{\mathbf{J}_{ij,1},\ldots,\mathbf{J}_{ij,S}\}
\end{equation}
with $(i,j)\in\{(1,1),(1,2),(2,2)\}$. Thus, we obtain
\begin{equation}\label{eq:FIM_global_block_form_app}
\mathbf{J}_{\boldsymbol{\xi}}=\begin{bmatrix}
\mathbf{J}_{11}&\mathbf{J}_{12}\\
\mathbf{J}_{12}^{\mathsf H}&\mathbf{J}_{22}
\end{bmatrix}.
\end{equation}

We further define the operator-form CRB as
\begin{equation}\label{eq:CRB_operator_XG}
\mathrm{CRB}(\mathbb{X},\mathbf{G},N_0)\triangleq\left(\mathbf{J}_{11}-\mathbf{J}_{12}\mathbf{J}_{22}^{-1}\mathbf{J}_{12}^{\mathsf H}\right)^{-1},
\end{equation}
where $\mathbf{J}_{11}$, $\mathbf{J}_{12}$, and $\mathbf{J}_{22}$ denote the FIM sub-blocks for all STs corresponding to the geometric parameters and the reflection-coefficient parameters, which can be constructed from $(\mathbb{X},\mathbf{G},N_0)$ via the stacked echo model and the Jacobians derived previously. Therefore, the CRB for jointly estimating the geometric parameters of the $s$-th ST is the $s$-th diagonal block of \eqref{eq:CRB_operator_XG}, written as
\begin{equation}\label{eq:CRB_final_app}
\mathrm{CRB}_s=\left(\mathbf{J}_{11,s}-\mathbf{J}_{12,s}\mathbf{J}_{22,s}^{-1}\mathbf{J}_{12,s}^{\mathsf H}\right)^{-1}.
\end{equation}
We see that \eqref{eq:Jlp_trace0}--\eqref{eq:CRB_final_app} are consistent with the compact formulations given in Section \ref{sec:II_system_C}.

\vspace{-10pt}
\bibliography{ref}
\bibliographystyle{IEEEtran}

\end{document}